\documentclass[prb, twocolumn, superscriptaddress, floatfix, tighten]{revtex4-2}
\usepackage{xcolor}
\usepackage{amsmath}
\usepackage{amssymb}
\usepackage{bm}      	
\usepackage{graphicx}
\usepackage[unicode=true,pdfusetitle,bookmarks=false,bookmarksnumbered=false,bookmarksopen=false,breaklinks=false,pdfborder={0 0 1},backref=false,colorlinks=true,urlcolor=blue,citecolor=blue,linkcolor=black]{hyperref}
\usepackage{subfigure}
\usepackage{textcomp}

\usepackage[english]{babel}
\usepackage{bbm,tabularx,subfigure}

\newcommand{\bsub}{\begin{subequations}}
\newcommand{\esub}{\end{subequations}}

\newcommand{\vex}[1]{\bm{\mathrm{#1}}}
\newcommand{\e}{\varepsilon}

\newcommand{\mathi}{\mathrm{i}}

\newcommand{\tmmathbf}[1]{\boldsymbol{#1}}
\newcommand{\tmop}[1]{\operatorname{#1}}

\newcommand{\NL}{\nonumber\\}
\newcommand{\da}{\downarrow}
\newcommand{\ua}{\uparrow}

\newcommand{\pupsf}[1]{{\scriptscriptstyle{({\mathsf{#1}})}}}

\newcommand{\intl}[1]{\int\limits_{#1}}
\newcommand{\T}{\mathsf{T}}

\begin{document}

\title{Topological surface-state destruction via trivializing proximity effect:\\
Lattice localization despite continuum criticality} 
\def\rice{Department of Physics and Astronomy, Rice University, Houston, Texas
77005, USA}
\def\rcqm{Rice Center for Quantum Materials, Rice University, Houston, Texas
77005, USA}
\author{Arthur Niwazuki}\affiliation{\rice}
\author{Matthew S. Foster}\affiliation{\rice}\affiliation{\rcqm}
\date{\today}

\begin{abstract}
In a significant conceptual revision to the tenfold classification scheme for topological insulators and superconductors,
it was recently demonstrated that most three-dimensional (3D) classes are simultaneously ``localizable'' in two distinct, but intricately connected ways: 
(1) There is no obstruction to Wannier localization of \emph{all} bulk eigenstates, and 
(2) \emph{Almost all} surface states can be Anderson localized by arbitrarily weak symmetry-preserving quenched disorder. Here we consider the localizable class CI in 3D, and numerically investigate the stability of surface states. We demonstrate that surface states of a bulk class-CI topological lattice model are fragile in that they can be Anderson localized by the combination of weak quenched randomness and hybridization with 
an additional trivial 2D band (a trivializing proximity effect, TPE). With the TPE, stronger disorder is more destructive to the surface states of the bulk lattice model.
By contrast, without additional bands the surface states remain extended, exhibiting robust spectrum-wide quantum criticality. We also investigate the fragility of surface states in effective 2D class-CI continuum Dirac theories, including the chiral limit of the Bistritzer-MacDonald model for twisted bilayer graphene. Although the continuum models exhibit signs of Anderson localization near gap edges for weak disorder, stronger disorder instead appears to \emph{heal} the surface, restoring criticality whilst filling in spectral energy gaps. Our results provide further evidence that effective continuum field theories fail to capture key aspects of surface-state physics in localizable topological phases.
\end{abstract}

\maketitle

\section{Introduction}

A central dichotomy in topological systems \cite{BernevigBook,hasan2010colloquium,QiZhang2011,AndoFu2015}
is defining bulk invariants on the lattice while describing protected boundary modes with continuum field theory \cite{AltlandBook}. 
Recent developments have called into question the general applicability of the latter. For the three-dimensional (3D) topological class AIII, 2D continuum theories fail to reflect the intrinsic fragility of most surface states to Anderson localization \cite{UFO24}.

Here we investigate the stability of surface states in another 3D topological class CI, using bulk lattice and surface continuum Dirac models. Our main results are summarized in Fig.~\ref{fig:theta} and Sec.~\ref{sec:resultSum}. We find that surface states of the lattice model can be localized by the simultaneous introduction of disorder and coupling to trivial 2D degrees of freedom (all whilst preserving the symmetry class), what we call the trivializing proximity effect (TPE). By contrast, continuum models exhibit Anderson localization only for weak disorder, and ``heal'' when the disorder strength is increased.
Localizability in classes AIII and CI \cite{UFO24} is reviewed in below in Secs.~\ref{sec:loctop}--\ref{sec:SWQC}, where we also introduce the models studied in this work.

\subsection{Localizable topology \label{sec:loctop}}

One of the main attractions of topological materials \cite{BernevigBook,hasan2010colloquium,QiZhang2011,AndoFu2015}
is the robust conducting edge or surface states that appear at the boundary of an otherwise incompressible bulk. 
The boundary states are anomalous in that they represent ``half'' of a normal $(d - 1)$-D conducting system (with the other half
transplanted to the opposite side of the $d$-dimensional bulk). An inferred physical consequence of this anomaly is that edge
and surface states are expected to be topologically protected against all manner of weak symmetry-preserving perturbations,
including quenched disorder and interactions. This attribute is famously responsible for the precise quantization of the
edge-state mediated Hall conductance in the integer quantum Hall effect \cite{AltlandBook}. 

This canonical story was however recently upended in Ref.~\cite{UFO24}, which argued that the standard ten-fold classification
scheme for topological insulators and superconductors \cite{Schnyder08,Kitaev09,Ryu10} was incomplete. In particular,
this work established that topological classes in one, two, and three dimensions (3D) must be further subdivided into two distinct categories: non-localizable and localizable classes. Different from the quantum Hall effect and the usual $\mathbb{Z}_2$ topological insulators, localizable classes
exhibit a host of unconventional features: 
(1) \emph{all} bulk states can be Wannier localized, 
(2) surface states can be \emph{detached} from the bulk in energy (via a direct or indirect gap) 
with a suitable symmetry-preserving surface perturbation, and 
(3) \emph{almost all} surface states can be Anderson localized by arbitrarily weak disorder. 
\emph{Zero-energy} surface states remain robustly protected \cite{Essin15,Schulz-BaldesBook}.

Non-localizable classes by contrast are guaranteed to possess an obstruction to exponential Wannier localization in the bulk. 
Edge or surface states are protected from Anderson localization by a spectral flow principle connecting the bulk and boundary spectra. 
This spectral-flow principle, which does not exist for localizable phases, 
is responsible for the precise quantization of the quantum Hall conductance involving current flow 
\emph{through the bulk} in the Corbino geometry \cite{Laughlin81,Halperin82}.

All one-dimensional (two-dimensional) topological phases are localizable (non-localizable), while three of five of the classes in 3D are localizable \cite{DIIInote}.
The refined classification scheme has since been extended to all dimensions \cite{Lapierre24,KawabataPRL25,KawabataPRB25}, with interesting connections established between detachable boundary states and the classification of non-Hermitian topology \cite{KawabataPRL25,KawabataPRB25}.

In this paper, we consider the localizable class CI, which describes time-reversal invariant, spin SU(2)-symmetric 3D topological superconductors \cite{Ryu10}. Employing a bulk three-dimensional lattice model \cite{schnyder2010andreev}, we use kernel-polynomial-method (KPM) numerics
\cite{weisse2006kernel,guan2021dual,Zhang25}
to demonstrate that surface states away from zero energy can be Anderson localized with weak disorder. We show that localization becomes possible when the surface is coupled to an additional 2D trivial flat band 
(in such a way that the composite system preserves class-CI symmetries), consistent with the expectations articulated in Ref.~\cite{UFO24}. In the latter, it was predicted that coupling trivial degrees of freedom to the surface can open up a spectral gap between bulk and surface states in localizable phases; the detached surface states at nonzero energy are then vulnerable to Anderson localization. We call this a trivializing proximity effect (TPE) that demonstrates the fragility of surface states in a localizable phase. By contrast, without the TPE we observe robust criticality of the surface states in the presence of disorder, a phenomenon known as ``spectrum-wide quantum criticality'' \cite{Ghorashi18,Sbierski20,ghorashi2020criticality,Karcher21,UFO24}, discussed more below.

\subsection{2D Dirac surface states and moir\'e simulators}

Typical topological surface states take the form of 2D massless Dirac fermions \cite{BernevigBook,hasan2010colloquium,QiZhang2011}. 
Although topological boundary Dirac fermions cannot be exactly realized in two-dimensional
crystals with local hopping (a consequence of the parity anomaly \cite{Redlich84,Semenoff84,Haldane88}), 
they can be \emph{effectively simulated} in 
such systems when intervalley coupling can be neglected. Hence, a single $K$ valley 
in graphene 
\cite{NetoGeim2009,DasSarmaRossi2010,DasSarmaRossi2011}
(neglecting spin) simulates the surface of a $\mathbb{Z}_2$ topological insulator,
and each nodal quasiparticle valley in the $d$-wave superconducting state of the cuprates 
\cite{Altland2002}
is approximately 
equivalent to the surface of $^3$He-$B$ \cite{ghorashi2020criticality}.
Bistritzer-MacDonald (BM) models for bilayer and trilayer moir\'e graphene 
\cite{Santos07,BM11,de_Gail11,MacDonald19,Tarnopolosky19}
in the chiral limit 
simulate class-CI and class-AIII surface states, respectively \cite{Zhang25,moiresimnote}.
Unlike the boundary modes of a topological bulk, none of these simulators are truly robust.
For example, 
scalar potential disorder is enough to destroy the analogy between 
moir\'e graphene and topological superconductor surface states \cite{footnoteSPDirt}. 

In this work we also study the effects of chiral-symmetric (``twist'') disorder 
\cite{wilson2020disorder,Padhi20,shavit2023strain,nakatsuji2022moire,guerrero2025disorderinduced,sanjuanciepielewski2025transport,Queiroz25,
uri2020mapping,kapfer2023programming}
on the chiral BM model for twisted bilayer graphene (TBLG). This model is effectively topological
due to the chiral and particle-hole symmetries needed for class CI, but also because
the $AB$-interlayer tunneling that couples the $K$ valleys between the two layers corresponds to a special
realization of this class.  
The advent of band gaps 
at the magic angle (e.g.) already demonstrates the absence of a spectral flow principle
for this type of (effective) topological surface theory: in this case, 
we do not need additional degrees of freedom beyond the moir\'e potential
to detach the ``surface'' and ``bulk'' states. Different from our results for
the topological lattice model, however, we find that increasing 
disorder whilst preserving class-CI symmetries
\emph{suppresses} localization effects.

\subsection{Spectrum-wide quantum criticality versus surface-state localizability \label{sec:SWQC}}

A special feature of topological superconductor surface-state Dirac models is their
exact solvability at zero energy in the presence of disorder
\cite{Nersesyan1994,Ludwig1994,Tsvelik1995,Mudry1996,Caux1996,Bhaseen2001},
which establishes the protection of these states from Anderson localization \cite{Evers08}.
The protection of \emph{finite-energy} surface states (i.e., those away from the surface Dirac point)
was only investigated recently. In numerical studies that employed effective continuum Dirac models
(defined in momentum space to avoid fermion doubling), it was found that finite-energy states
also avoid Anderson localization. Instead, finite-energy Dirac surface-states of class CI and AIII 
topological phases become quantum critical, with universal spatial fluctuations 
respectively captured by the spin- and integer-quantum Hall plateau transitions
\cite{Ghorashi18,Sbierski20,ghorashi2020criticality,Karcher21,UFO24}.
This phenomenon was
dubbed spectrum-wide quantum criticality (SWQC).

Ref.~\cite{UFO24} questioned the ubiquity of SWQC. Focusing on the localizable class AIII, it was demonstrated that a certain symmetry-preserving, homogeneous modification at the surface of a certain bulk topological lattice model 
(1) allows the surface states to be detached from the bulk (via an indirect or direct energy gap) and 
(2) enables Anderson localization of all surface states with energy $E \neq 0$.
In addition, the perturbing lattice potential is \emph{completely absent} from the effective 2D continuum surface Dirac theory, because it projects to zero in that subspace. 
In the case of class AIII, the mechanism for surface-state detachment and Anderson localization was shown to be the injection of Berry curvature ``from the sky,'' i.e.\ from the surface-bulk spectral interface. 
The Berry curvature enables Anderson localization of finite-energy surface states, forming a topological Anderson insulator (as in the integer quantum Hall effect or a dirty Chern insulator). 
At the same time, it is possible to arrange for the surface to possess a certain statistical symmetry so that chiral edge states bounding different random surface domains percolate. This restores surface SWQC, and explains why all finite-energy surface states then exhibit universal integer-quantum-Hall plateau-transition statistics \cite{Sbierski20,UFO24}.

Our focus here is class CI, which possesses a $T^2 = +1$  time-reversal symmetry \cite{Schnyder08,Ryu10}.  
In this case, Berry curvature cannot be the key to disorder-mediated surface-state annihilation. We instead exploit the TPE to demonstrate localizability in a class-CI topological bulk lattice model. 

In addition to bulk lattice and 2D continuum BM models, we consider the coupling of the dirty 2D continuum CI Dirac model from Ref.~\cite{Ghorashi18} to a continuum flat band. We find that although the coupling to trivial degrees of freedom does allow Anderson localization of some surface states with weak disorder, stronger disorder appears to \emph{heal} the continuum model, restoring SWQC. As with the case of class AIII in Ref.~\cite{UFO24}, we take this as further evidence that continuum surface field theories (e.g., 2D dirty Dirac equations) are \emph{fundamentally inadequate} for capturing the surface physics of localizable topological phases.

\begin{figure}[bh!]
    \centering
    \includegraphics[width=0.48\textwidth]{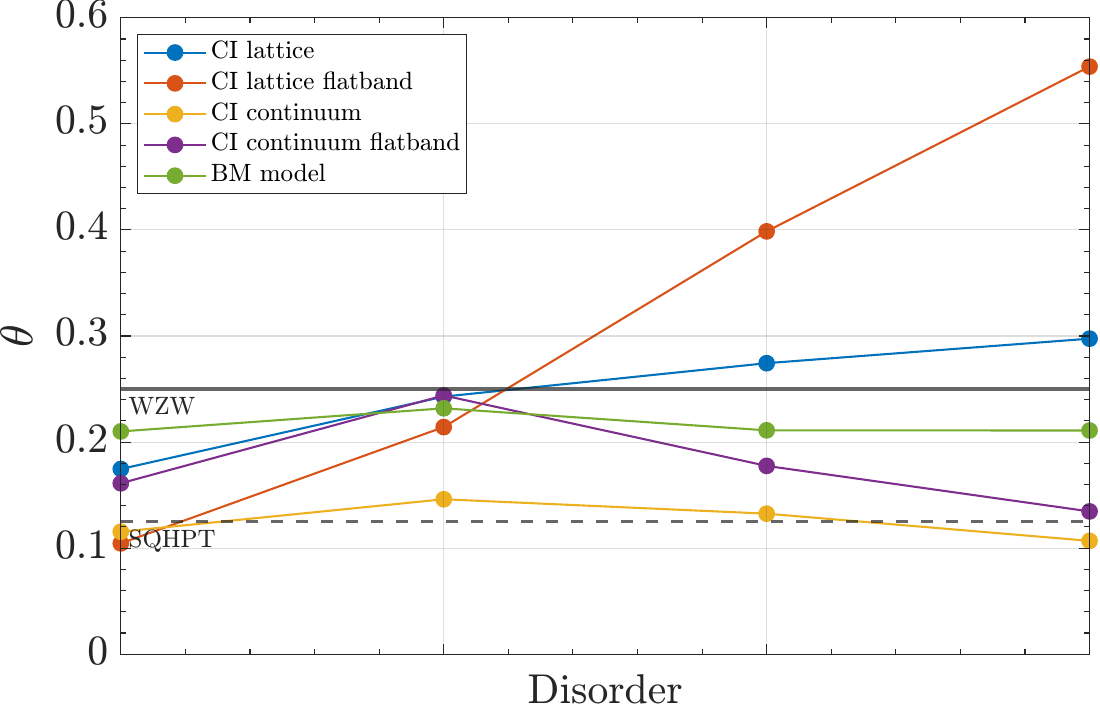}
    \caption{\label{fig:theta}
    Contrasting the behavior of lattice and continuum models of class-CI topological surface states,
    subject to disorder and a trivializing proximity effect (TPE). 
    As a proxy for wave function criticality or Anderson localization,
    we plot the numerically computed curvature parameter $\theta$ for the multifractal spectrum (see text),
    averaged over a certain finite energy window, versus the effective disorder strength. 
    Larger values of $\theta$ indicate stronger spatial rarefaction, tending towards Anderson localization.
    Results are shown for 5 different models: 
    2D surface states of a 3D class-CI lattice model without and with coupling to a trivial class-CI 2D flat band,
    the same for states of a 2D continuum Dirac theory of the surface,
    and the continuum Bistritzer-MacDonald (BM) model for twisted bilayer graphene in the chiral limit, subject
    to chiral disorder. The BM model is tuned to the first magic angle; the band gaps play a similar role as the coupling to a trivial flat band for the surface states.
    The lines marked ``WZW'' and ``SQHPT'' refer to predictions for zero- and finite-energy \emph{critical} surface states, in the so-called spectrum-wide quantum criticality (SWQC) scenario 
    \cite{Ghorashi18,Karcher21,UFO24}.
    The main takeaways are
    (1) Increasing disorder localizes lattice surface states (away from zero energy) 
    with the TPE, but not significantly without it,
    and
    (2) In the 2D continuum theories, increasing disorder has either no effect 
    (states remain critical) 
    or \emph{suppresses} an initial trend towards localization for weak disorder. Disorder strengths are selected relative to each model and can only be qualitatively compared across models;
    details for these and the selected energy windows are provided in Appendix \ref{sec:fig1params}.
    }
\end{figure}

\subsection{Summary sketch of the main results \label{sec:resultSum}}

We preview our main results in Fig.~\ref{fig:theta}. Here we plot a numerically computed parameter $\theta$ for several different models, as a function of disorder strength. The parameter $\theta$ is (half) the curvature of the so-called \emph{multifractal spectrum} \cite{Evers08}, used here to measure the extended or localized nature of 2D surface-state wave functions
[see Eqs.~(\ref{IPR})--(\ref{thetaDef}), below, for a precise definition].
Larger values of $\theta$ indicate a tendency towards Anderson localization. 
Presented results obtain via fixed realizations of disorder, averaged over large energy windows of surface-state spectra (see Appendix~\ref{sec:fig1params} for details).
Two horizontal lines in this figure mark special values of $\theta$. 
The line denoted ``WZW'' has $\theta = 1/4$, which is the exact result for the class-CI, quantum-critical zero-energy continuum Dirac surface state in the presence of symmetry-preserving disorder 
\cite{Mudry1996,Caux1996}.
The line marked ``SQHPT'' 
has $\theta = 1/8$, and 
refers to the spin quantum Hall plateau transition; this is the approximate known result from extensive numerics for the plateau transition in class C \cite{Evers08}. 
Previous studies of 2D continuum Dirac theories exhibited SWQC, with SQHPT statistics for finite-energy states \cite{Ghorashi18,Karcher21}. 

Results are exhibited for 5 different models in Fig.~\ref{fig:theta}:
(1) the surface states of a bulk lattice model with disorder,
(2) the same as (1), but incorporating in addition coupling to a 2D trivial flat band (such that the composite system also resides in class CI),
(3) the continuum 2D Dirac model for dirty surface states from Ref.~\cite{Ghorashi18},
(4) the same as (3), but coupled to a 2D trivial flat band, and
(5) the chiral BM model for TBLG, subject to class-CI preserving disorder. 
Cases (1) and (3) show robust wave function criticality, with
the continuum case (3) matching well the SQHPT prediction. 
Case (2) demonstrates the TPE: finite-energy surface states localize more strongly with increasing disorder when coupled to a trivial 2D band of additional states. 
For cases (4) and (5), weak disorder produces localized states near band edges, as we demonstrate in Sec.~\ref{sec:result}.
Somewhat surprisingly, however, cases (4) and (5) show that stronger disorder appears to ``heal'' the surface, restoring SWQC and \emph{suppressing} the tendency towards localization. 

It is worth noting that the lattice-model results effectively describe surface states of an order-of-magnitude smaller surface, compared to the 2D continuum-model results (owing to computational limitations).
This is because the surface states occupy a relatively small fraction of the surface Brillouin zone for the bulk lattice model in the slab geometry.
Larger relative finite-size corrections are the likely reason why the lattice results without the TPE match better the WZW, rather than SQHPT prediction. Details of the parameters used to produce all data in Fig.~\ref{fig:theta} are provided in Appendix~\ref{sec:fig1params}.

\subsection{Outline}

The rest of this paper is organized as follows. 
In Sec.~\ref{sec:Models}, we introduce the bulk lattice and 2D continuum models employed in this work, and summarize key properties. 
This includes the effects of hybridizing surface states with trivial 2D bands in the absence of disorder, 
the incorporation of chiral-symmetric disorder,
and comparison to the BM model for TBLG. 
In Sec.~\ref{sec:result} we present numerical results for the topological robustness of surface states in the presence of disorder and the TPE, using KPM and exact diagonalization for the five models highlighted in Fig.~\ref{fig:theta}. We conclude with a list of key unanswered questions in Sec.~\ref{sec:oq}.


\section{Models \label{sec:Models}}

\subsection{CI bulk lattice model}

We consider the 2-orbit-per-site cubic lattice model 
for a mean-field class-CI topological superconductor introduced in Ref.~\cite{schnyder2010andreev}. 
The quasiparticle Hamiltonian is
\begin{gather}
  H_{\mathsf{L}}^\pupsf{3D} 
  = 
  \intl{\tmop{BZ}} 
    \frac{d^3 \vex{k}}{(2 \pi)^3} \, 
  \eta^{\dag} (\tmmathbf{k}) \, h_{\mathsf{L}}^\pupsf{3D} (\tmmathbf{k}) \, \eta (\tmmathbf{k}),
\NL
  \eta (\tmmathbf{k}) 
  = 
  \left[ 
        a_{\ua} (\tmmathbf{k}), 
        b_{\ua} (\tmmathbf{k}), 
        a^{\dag}_{\da} (-\tmmathbf{k}), 
        b_{\da}^{\dag} (-\tmmathbf{k}) 
    \right]^T,
\end{gather}
where $\eta(\vex{k})$ denotes the Nambu spinor,
$\vex{k} = \{k_x,k_y,k_z\}$ runs over the 3D cubic Brillouin zone (BZ),
and the Bogoliubov-de Gennes Hamiltonian is 
\begin{align}
  h_{\mathsf{L}}^\pupsf{3D}  (\tmmathbf{k}) 
  =&\, 
  \left(- \cos k_x - \cos k_y - \cos k_z - \mu\right) 
  \tau_3 
\NL
  &\,
  +
  \Delta_1  
  \left(\cos k_x - \cos k_y\right) \tau_1 \, \sigma_3 
\NL
  &\,
  +
  \Delta_2 
  \,
  \sin k_x \, \sin k_y \,
  \tau_1 \, \sigma_1 
  - 
  \Delta_2 
  \,
  \sin k_z 
  \,
  \tau_1 \, \sigma_2.\!\!
  \label{eq:hamiltonian}
\end{align}
Here $\{\tau_i\}$ and $\{\sigma_j\}$ denote Pauli matrices acting
respectively on particle-hole and orbital spaces,
$\mu$ is the chemical potential,
$\Delta_1$ encodes the strength of $d$-wave pairing in the $k_{x,y}$ plane,
and 
$\Delta_2$ is an interorbital pairing. 
We have set the lattice spacing to one and the normal-state hopping amplitude $t = -1$;
then this model is topological with winding number $\nu=2$ when $-3< \mu< -1$.

\subsubsection{Surface states}

\begin{figure}[t!]
  \centering
  \subfigure[]{
    \includegraphics[width=0.30\textwidth]{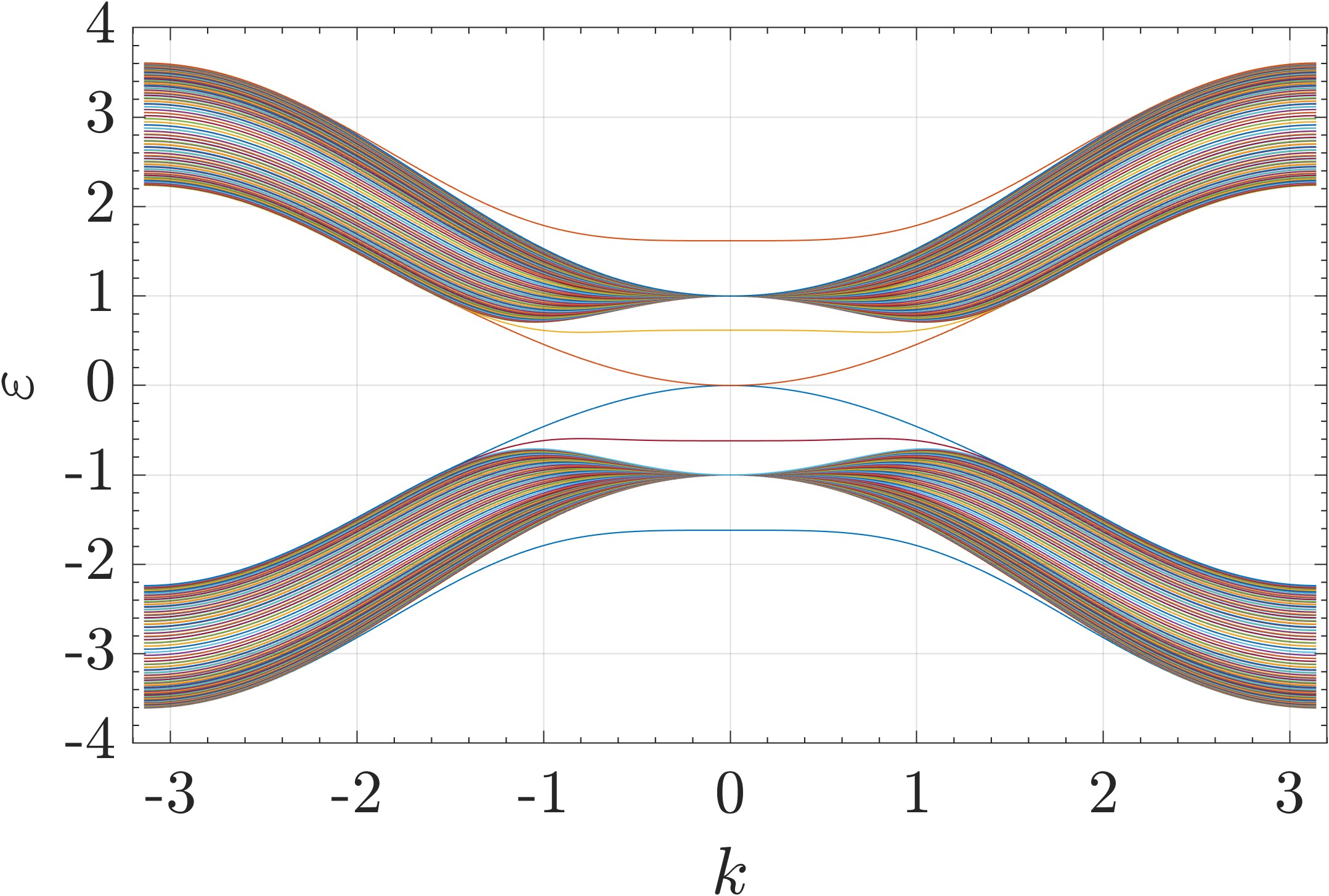}
  }
  \subfigure[]{
    \includegraphics[width=0.30\textwidth]{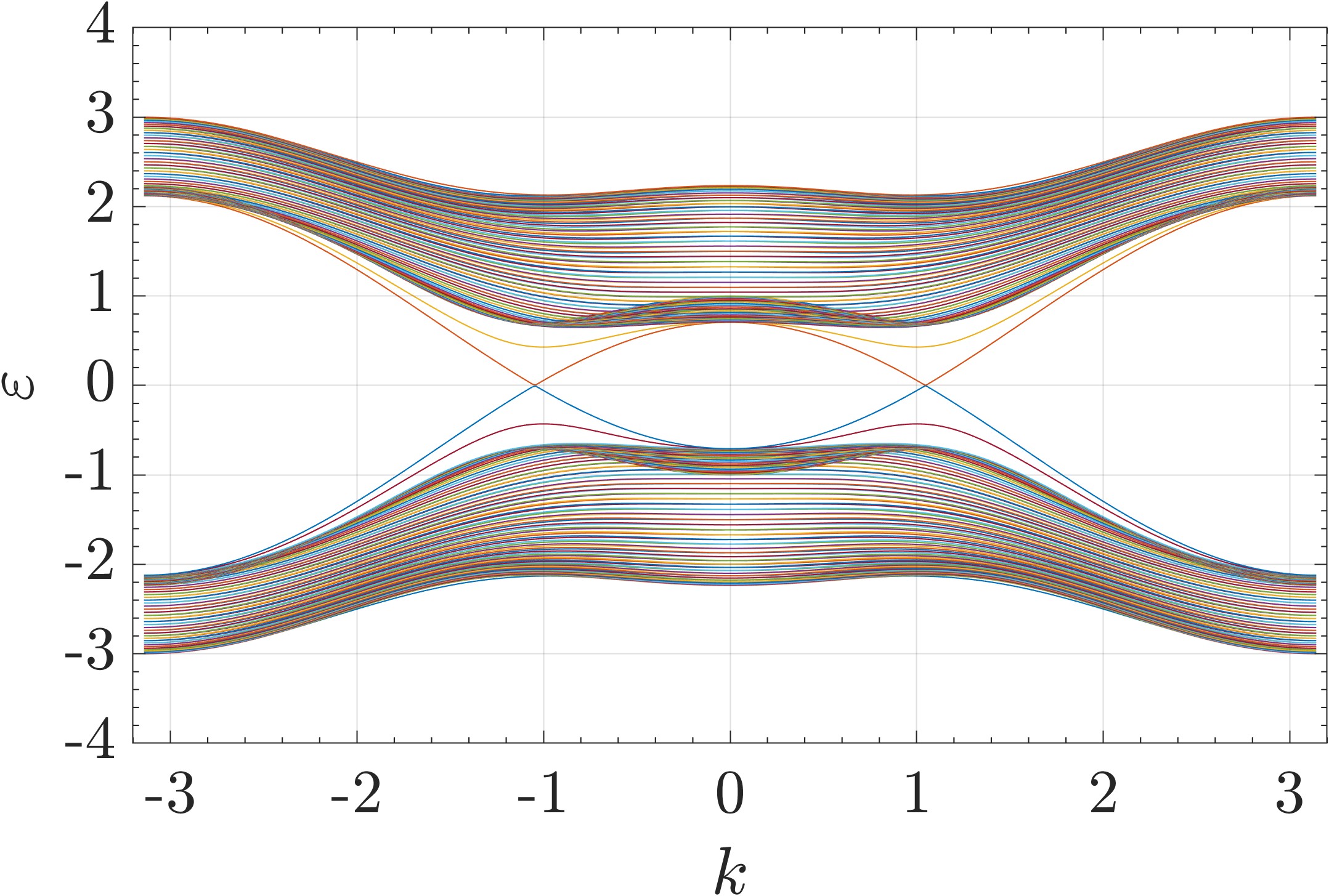}
  }
  \caption{\label{fig:ci_surf_spec}
  Slab spectrum of the cubic-lattice class-CI topological superconductor model defined by the 
  lattice Hamiltonian in Eq.~(\ref{eq:hamiltonian}) with winding number $|\nu| = 2$.
  (a) $z$-cut, $k_y = 0$ line. The surface states form a quadratic touching, analogous to a single valley of Bernal bilayer graphene.  
  (b) $x$-cut, $k_z = 0$ line. The surface states appear as two Dirac cones shifted away from the surface $\Gamma$ point. 
  The parameters for these plots are: $\mu = - 2$, $\Delta_1 = 1$, and $\Delta_2 = 1$.
  In addition, a surface Chern mass $m_c = 1$ is used to gap out one of the two slab surfaces in each plot,
  so as to energetically isolate the low-energy surface states on the ``active'' surface. 
  The slab thickness is $N_{z / x} = 60$.}
\end{figure}

Topological superconductors with $|\nu|=2$ 
typically possess two surface Dirac cones (with the Dirac points appearing at \emph{exactly} zero energy,
defined relative to the Fermi energy  in the paired bulk). 
For the lattice model in Eq.~(\ref{eq:hamiltonian}),
this is the case for an $x$-cut surface (parallel to the $yz$ plane).
By contrast, for the $z$-cut surface the model exhibits 
2D surface states with a quadratic touching, analogous to a single valley of 
Bernal-stacked bilayer graphene \cite{NetoGeim2009}.
The quadratic touching occurs at the $\Gamma$ point of the surface BZ, 
and has the low-energy continuum expansion
\begin{equation}
    h_{z}^\pupsf{2D}(k_x,k_y) = \frac{1}{M}  \left[\begin{array}{cc}
    0 & k^2\\
    \overline{k}^2 & 0
  \end{array}\right],
\end{equation}
where $k = k_x - \mathi \, k_y$ and $k$ is measured from the $\Gamma$ point. 
The spectra of the $z$-cut and $x$-cut surfaces are shown in Fig.~\ref{fig:ci_surf_spec}.

\subsubsection{Surface states with flat-band coupling}

\begin{figure}[b!]
  \centering
  \subfigure[]{
    \includegraphics[width=0.34\textwidth]{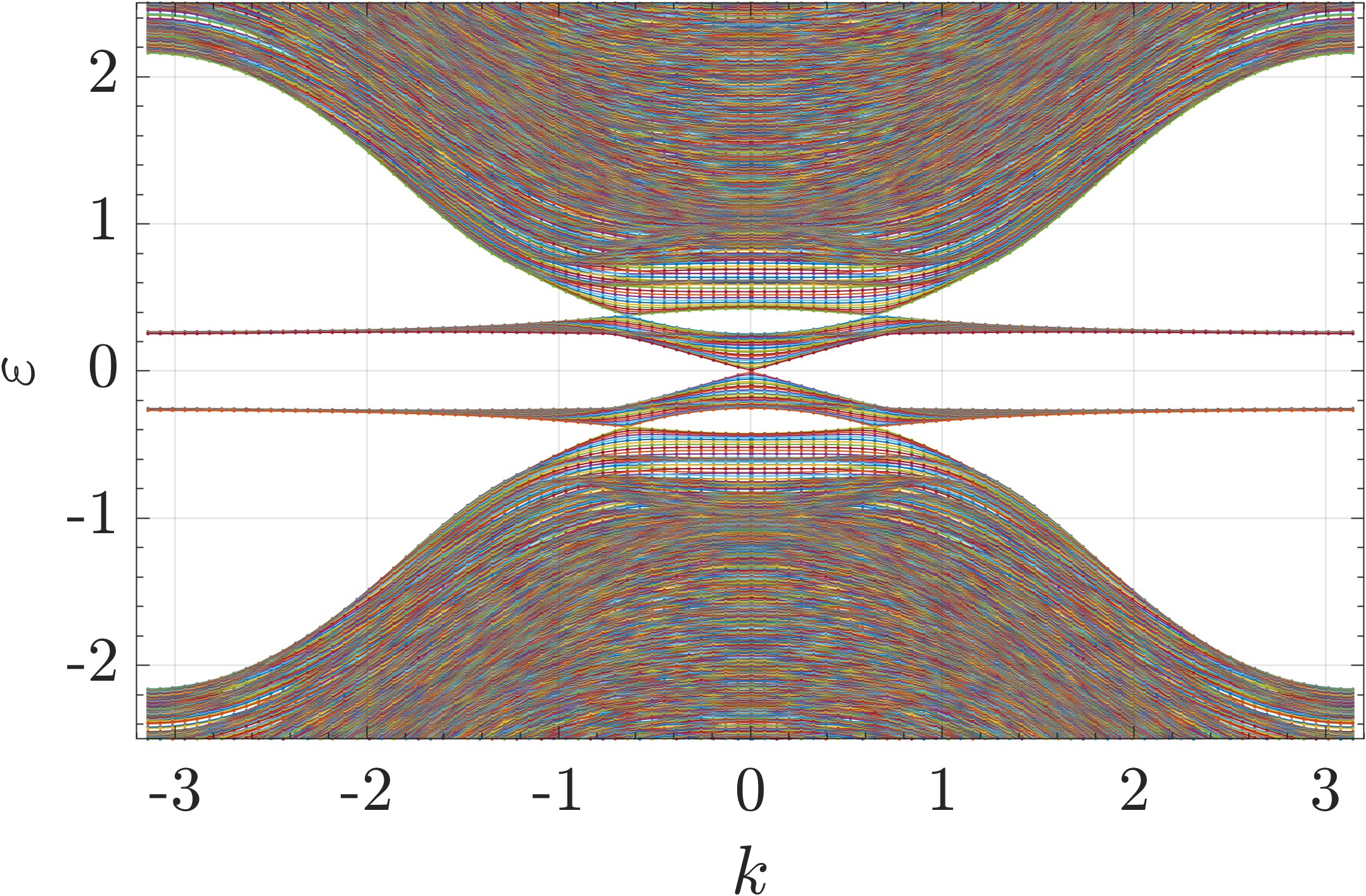}
  }
  \subfigure[]{
    \includegraphics[width=0.34\textwidth]{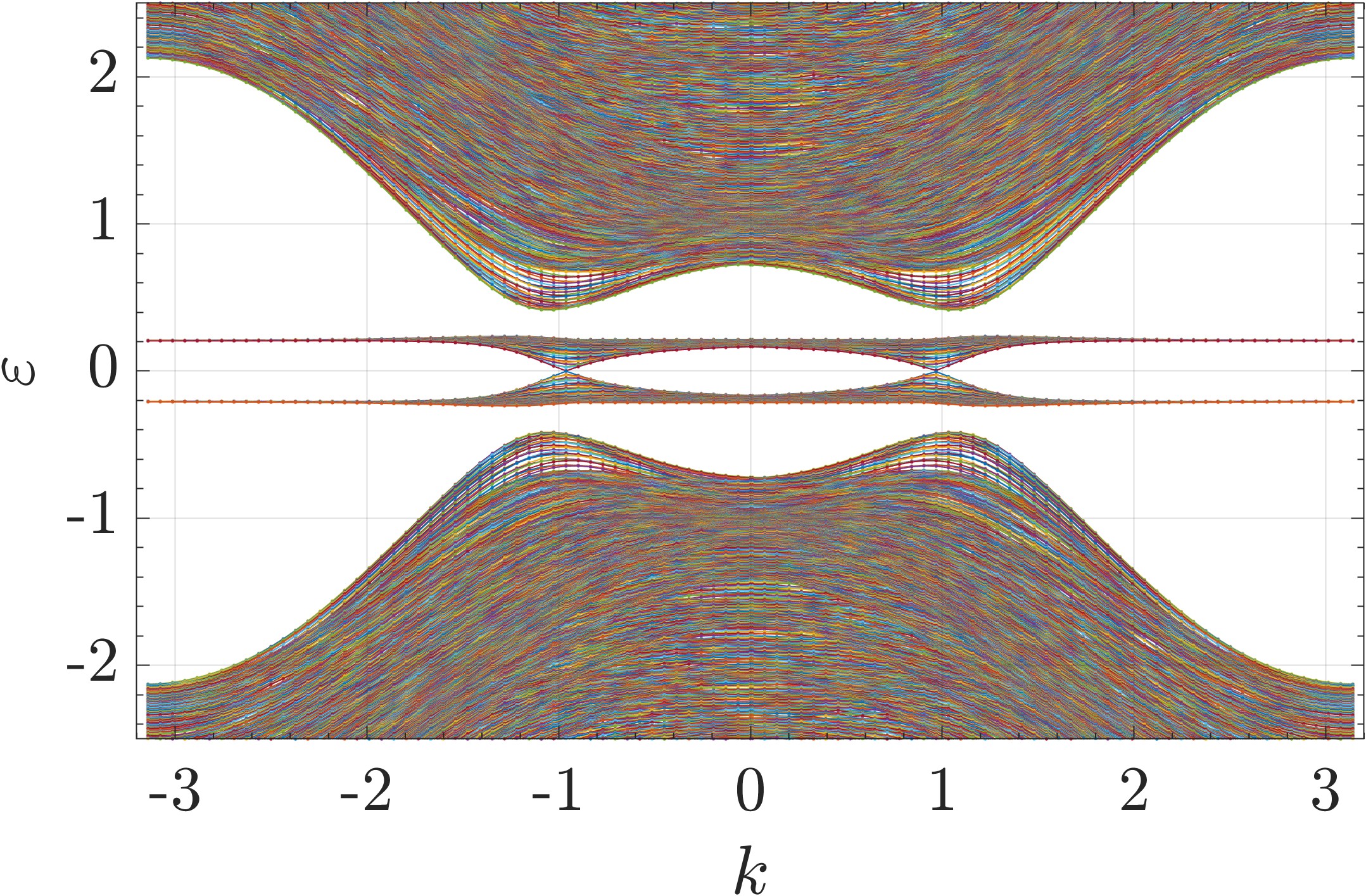}
  }
  \caption{\label{fig:lattice_flat_spec}
  Slab energy spectrum versus $k_y$ of the topological CI lattice model 
  (as in Fig.~\ref{fig:ci_surf_spec}), now coupled to a 2D flat band via Eq.~(\ref{eq:HamTPE}).
  (a) $z$-cut. (b) $x$-cut.
  The model parameters are 
  $\varepsilon_c = 0.2$, 
  $\gamma = 0.35$, 
  $\mu = - 2$, 
  $\Delta_1 = 1$,
  and
  $\Delta_2 = 1$.
  Again a Chern mass of size $m_c = 1$ is employed to gap out one of the two 
  slab surfaces. 
  (a) has $N_x = 100$, $N_z = 10$ and (b) is the same with $x \Leftrightarrow z$.  
  }
\end{figure}

\begin{figure}[t!]
  \centering
  \subfigure[]{
    \includegraphics[width=0.3\textwidth]{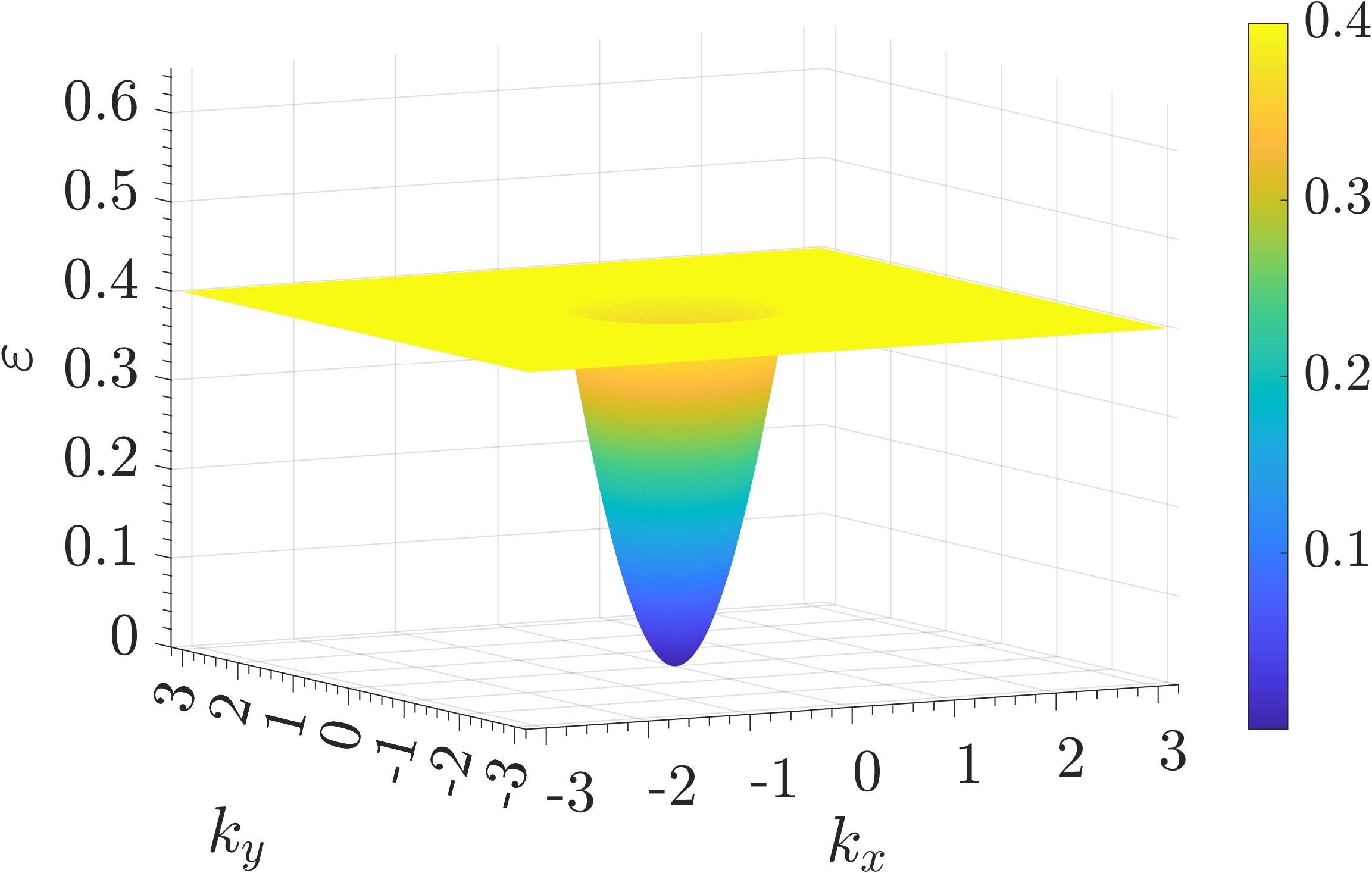}
  }
  \subfigure[]{
    \includegraphics[width=0.3\textwidth]{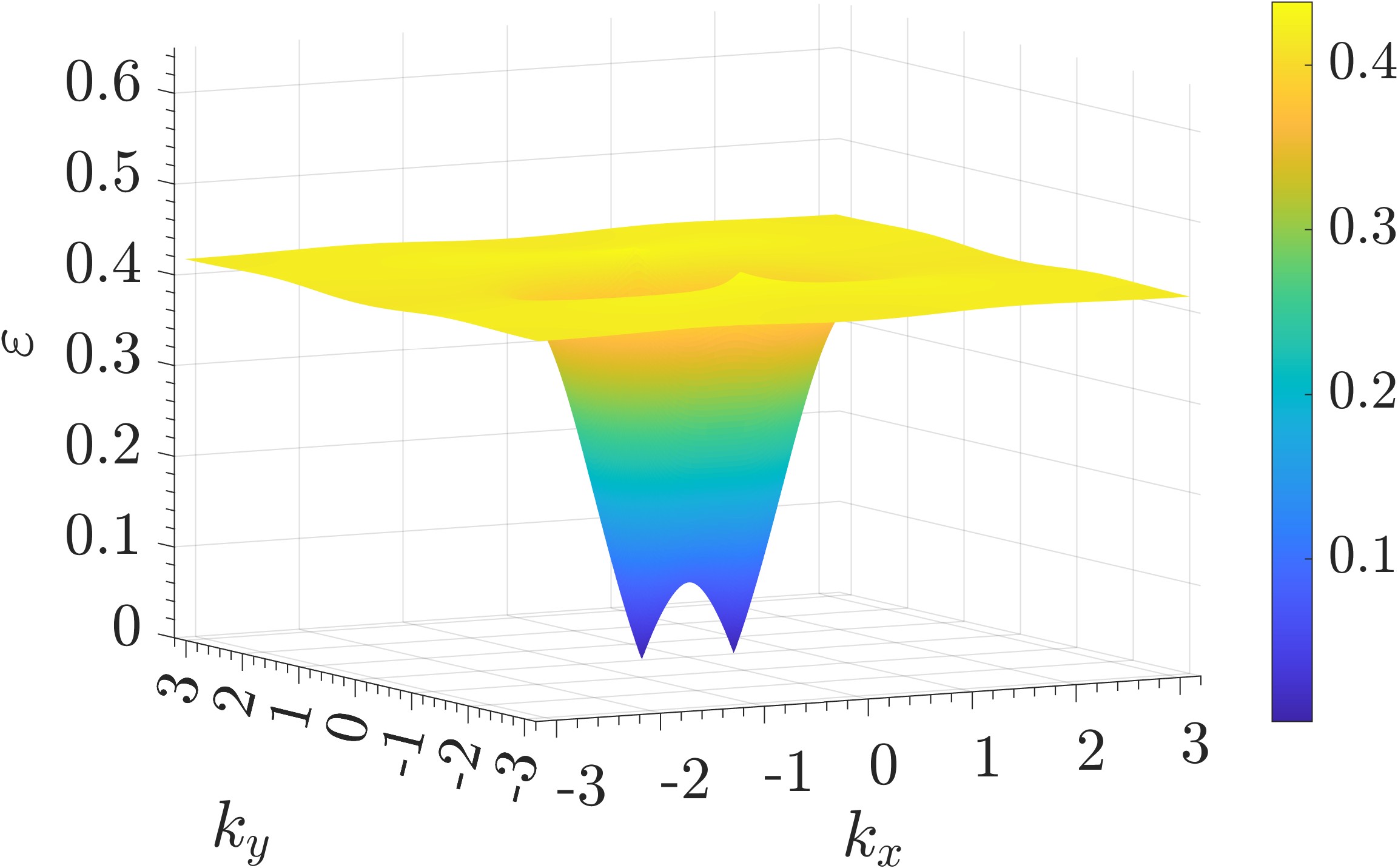}
  }
  \subfigure[]{
    \includegraphics[width=0.3\textwidth]{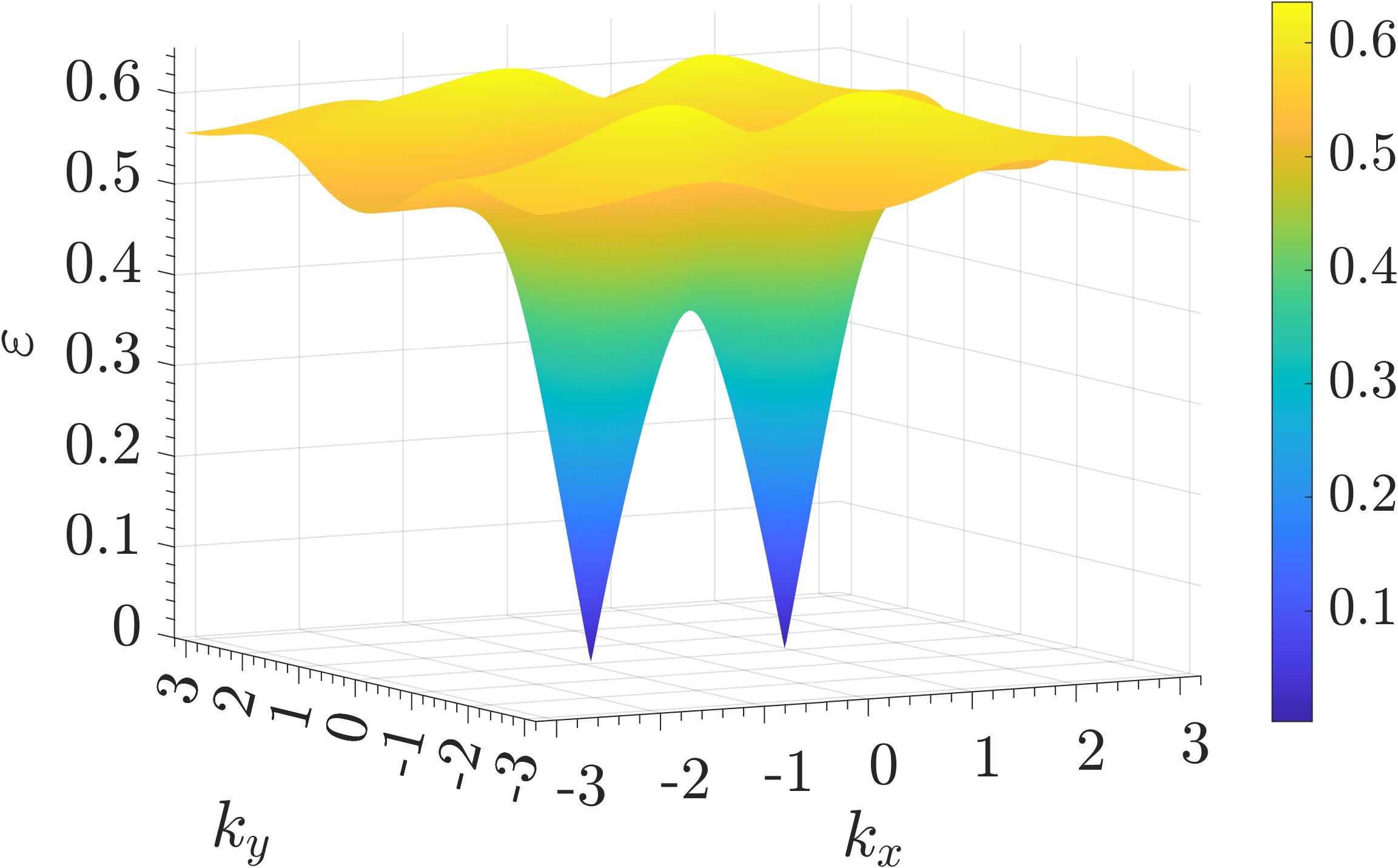}
  }
  \caption{\label{fig:lattice_z_sur}
    Splitting of the quadratic surface-state touching by proximity coupling. 
    We plot the lowest-energy band for the $z$-cut bulk CI topological lattice model hybridized
    with the trivial 2D flat band as in Fig.~\ref{fig:lattice_flat_spec}(a), for various hybridization
    strengths.   
    (a) $\gamma = 0$. (b) $\gamma = 0.2$. (c) $\gamma = 0.6$. 
    Other parameters are $\varepsilon_c = 0.4$, $\mu = - 2$, $\Delta_1 = 1$, 
    $\Delta_2 = 1$, $N_z = 20$. 
  }
\end{figure}

We will demonstrate that the surface states exhibited in Fig.~\ref{fig:ci_surf_spec}(a) become fragile (vulnerable to Anderson localization with \emph{weak} disorder) when these are coupled to an additional 2D flat-band lattice model.
Taking the topological bulk to have $z \geq 1$, we can couple it 
to a flat-band layer at $z = 0$ by adding the Hamiltonian
\begin{multline}\label{eq:HamTPE}
    \delta H_{\mathsf{TPE}}^\pupsf{3D} 
    =
    \frac{\varepsilon_c}{2}  
    \intl{\tmop{SBZ}}
    \frac{d^2 \vex{k}}{(2 \pi)^2}
    \,  
    \eta^{\dag}_{z = 0}(k_x,k_y) 
    \,
    \tau_3 
    \,
    \eta_{z = 0}(k_x,k_y) 
\\
    + 
    \gamma
    \left[
    \begin{aligned}
    &\,
        \sum_{\sigma} 
        \intl{\tmop{SBZ}}
        \frac{d^2 \vex{k}}{(2 \pi)^2}
        \,
        \eta^{\dag}_{z = 0}(k_x,k_y) 
        \,
        \tau_1 
        \,
        \eta_{z = 1,\sigma}(k_x,k_y)
    \\
    &\,    
        + \mathrm{H.c.},
    \end{aligned}
    \right]\!.\!\!
\end{multline}
Here $\eta_{z = 0}$ is a two-component Nambu spinor in particle-hole space
of the flat band, 
$\varepsilon_c$ sets the energy of the flat band quasiparticles (at $\pm \varepsilon_c$), 
and
$\gamma$ denotes the (purely anomalous) coupling strength between the 2D flat band and
the top layer of the topological superconductor bulk crystal. The integrals run over the entire surface Brillouin zone (SBZ). 

The combined bulk and flat-band system still resides within class CI. 
It has chiral (physical time-reversal \cite{Schnyder08}) symmetry
\begin{equation}
    \eta_{x y z} 
    \rightarrow M_S 
    \left(\eta_{x y z}^{\dag}\right)^\T, 
\quad
    \mathi \rightarrow - \mathi, 
\quad
    M_S = \tau_2,
\end{equation}
and 
$P^2 = -1$ particle-hole
(physical spin $\pi$-rotation around $\hat{x}$ \cite{Schnyder08})
symmetry
\begin{equation}
    \eta_{x y z} 
    \rightarrow 
    M_P 
    \left(\eta_{x y z}^{\dag}\right)^\T, 
\quad
    M_P = \tau_2.
\end{equation}

Coupling of the $z$-surface of the topological bulk to the 2D flat band is shown
in Fig.~\ref{fig:lattice_flat_spec}(a), whilst an analogous coupling to 
the $x$-cut surface is shown in panel (b) of this figure. 
In the absence of disorder, coupling with the flat band opens up a direct gap between the bulk 
and surface spectra for coupling to the $x$ surface. For coupling to the $z$-cut surface,
there is no gap introduced by the flat band. Nevertheless, we will find that this coupling is sufficient to make the surface states fragile to the effects of disorder. 

Similar to the regime of \emph{very small} (rigid) twists in TBLG \cite{de_Gail11}, coupling of the $z$-cut surface to the flat band splits the quadratic touching of the pristine surface states into two Dirac cones, as shown in 
Fig.~\ref{fig:lattice_z_sur}. Here we plot the spectrum for the lowest-energy slab band corresponding to the hybridization of the $z$-cut surface of the topological bulk with the 2D trivial flat band.


\subsection{CI continuum Dirac surface model}

Next we consider the 
2D continuum Dirac model for class-CI surface states used to demonstrate
SWQC in Ref.~\cite{Ghorashi18},
\begin{equation}\label{hx2D}
    h_{x}^\pupsf{2D}  
    =
    -
    \mathi
    \,
    \vex{\sigma} \cdot \vex{\nabla}
    +
    \tmmathbf{A}^j(\vex{r}) 
    \cdot
    \vex{\sigma} 
    \,
    \tau_j,
\end{equation}
where $\vex{\nabla}$ is the 2D gradient operator,
$\{\sigma_i\}$ and $\{\tau_j\}$ are Pauli matrices acting on the spinor and ``color'' spaces, 
and the repeated index $j$ is summed over $\{1,2,3\}$. 
Eq.~(\ref{hx2D}) is the effective 2-cone surface Dirac Hamiltonian for the $x$-cut topological bulk lattice model,
Fig.~\ref{fig:ci_surf_spec}(b), after setting the low-energy Fermi velocity of the cones to one.
We have in addition incorporated generic class-CI surface disorder,
which takes the form of the SU(2) non-abelian vector potential
$A^i_\mu(\vex{r})$ \cite{Schnyder08,footnoteSPDirt}, 
and which connects the two Dirac colors in exactly the same way as the moir\'e matrix potential couples together layers 
in the chiral-limit of the BM model for TBLG \cite{Santos07,BM11,Tarnopolosky19}.

\begin{figure}[t!]
  \centering
  \includegraphics[width=0.35\textwidth]{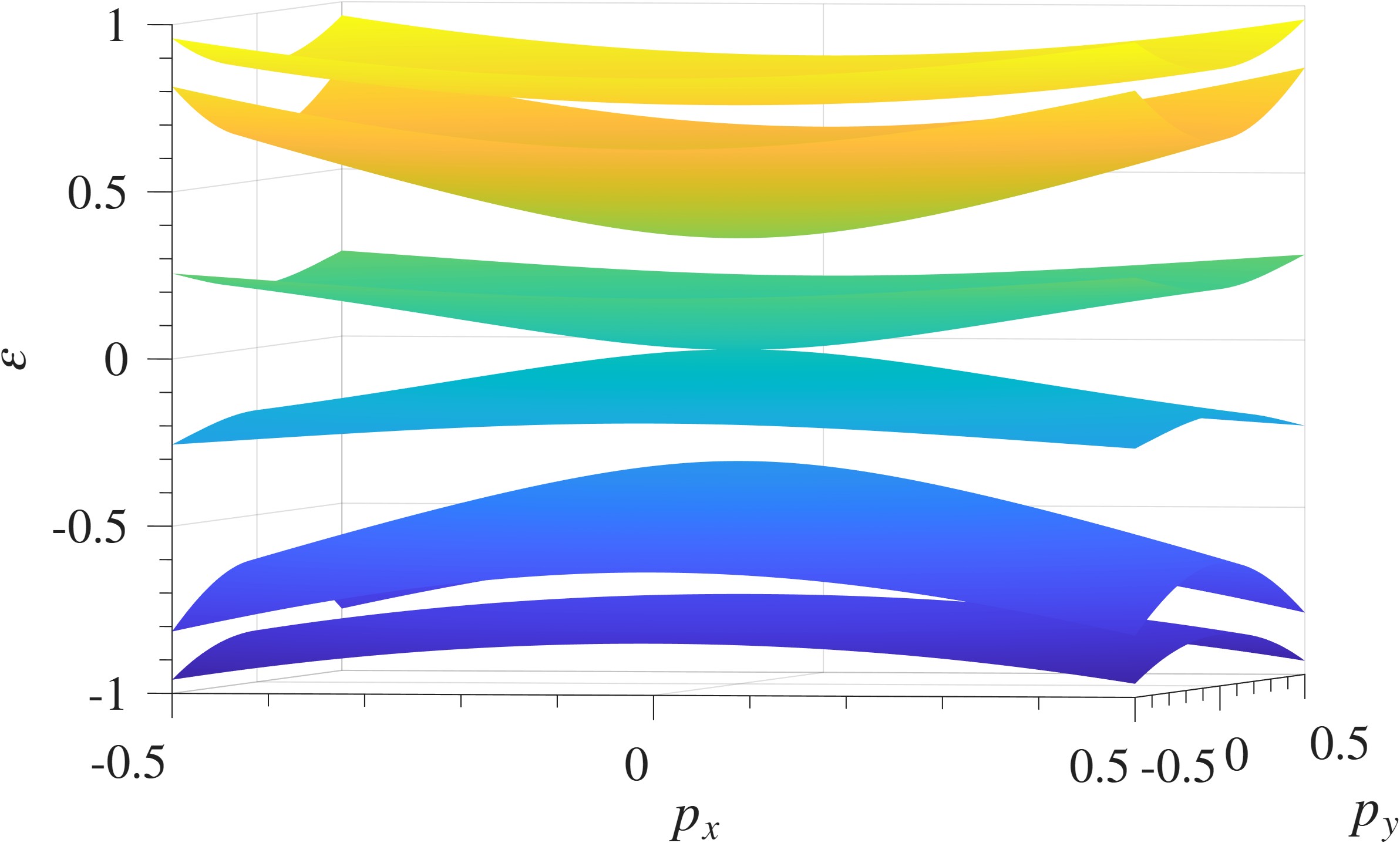}
  \caption{\label{fig:ci_continu_spec}Spectrum of the CI 2D continuum surface-Dirac model hybridized with a class-CI flat band,
  in the absence of disorder. 
    The parameters are $\varepsilon_c = 0.4$ and $\gamma_0 = 0.35$.}
\end{figure}

We can further couple this continuum surface theory to a continuum flat band.
As in Ref.~\cite{Ghorashi18}, we formulate the full model in momentum
space in order to avoid fermion doubling,
\begin{widetext}
\begin{align}\label{HDiracTPE}
    H_{\mathsf{TPE}}^\pupsf{2D} 
    =
    \sum_{\vex{p_1},\vex{p_2}} 
    \left[
    \begin{aligned}
    &\,
        \psi^{\dag}_{\vex{p_1}}  
            \left[
                    \vex{p_1} \cdot \vex{\alpha}
                    \,
                    \delta_{\vex{p_1}, \vex{p_2}} 
                    +
                    \vex{A}^j (\vex{p_1} - \vex{p_2}) 
                    \cdot 
                    \vex{\alpha}
                    \,
                    T_j
            \right] 
        \psi_{\vex{p_2}}  
        + 
        \phi_{\vex{p_1}}^{\dag} 
            \left[
                \varepsilon_c
                \,
                \delta_{\vex{p_1},\vex{p_2}} 
                + 
                A_c(\vex{p_1} - \vex{p_2})
            \right] 
            \tau_1 
        \phi_{\vex{p_2}} 
    \\
    &   + 
        \sum_{\sigma = 1}^2 
        \left\{
            \phi_{\vex{p_1}}^{\dag} 
            \left[\gamma_0 \,   \delta_{\vex{p_1}, \vex{p_2}} + \gamma_1(\vex{p_1} -\vex{p_2}) \right] 
            \tau_1  \, \psi_{\sigma, \vex{p_2}}
            +  
            \phi_{\vex{p_1}}^{\dag} 
            \left[\gamma_0 \,   \delta_{\vex{p_1}, \vex{p_2}} + \gamma_2(\vex{p_1} -\vex{p_2}) \right] 
            \tau_2 \, \psi_{\sigma, \vex{p_2}} 
            + 
            \mathrm{H.c.}
        \right\} 
    \end{aligned}
    \right],
\end{align}
\end{widetext}
where after a basis change the Dirac kinetic matrices and SU(2) color generators are 
\begin{align}
  \{ \tmmathbf{\alpha} \} = \{ \tau_2 \, \sigma_2, - \tau_1 \, \sigma_2 \},
  \quad
  \{ T_j \} = \{ - \tau_3 \, \sigma_3, \sigma_2, \tau_3 \, \sigma_1 \}.
\end{align}
In Eq.~(\ref{HDiracTPE}), $\psi \rightarrow \psi_{\tau,\sigma}$ denotes the 4-component Dirac surface state,
while $\phi \rightarrow \phi_\tau$ is a 2-component trivial flat-band Nambu spinor. 
The flat band has energy $\varepsilon_c$ with a random, zero-average component $A_c(\vex{r})$. 
The (purely anomalous) couplings between the Dirac surface and flat-band states are given by 
the average parameter and random, zero-mean functions $\gamma_0$ and $\gamma_{1,2}(\vex{r})$,
respectively.

The composite system in Eq.~(\ref{HDiracTPE}) is invariant under chiral and $P^2 = -1$ particle-hole transformations required by class CI. These are encoded in the symmetry transformations 
\begin{equation}
    \psi 
    \rightarrow M_S 
    \left(\psi^{\dag}\right)^\T\!\!, 
\quad
    \phi 
    \rightarrow M_S 
    \left(\phi^{\dag}\right)^\T\!\!, 
\quad
    \mathi \rightarrow - \mathi, 
\quad
    M_S = \tau_3,
\end{equation}
and 
\begin{equation}
    \psi 
    \rightarrow M_P 
    \left(\psi^{\dag}\right)^\T\!\!, 
\quad
    \phi 
    \rightarrow M_P 
    \left(\phi^{\dag}\right)^\T\!\!, 
\quad
    M_P = \tau_2.
\end{equation}
The particle-hole condition forces $\gamma_{0}$ and $\gamma_{1,2}(\vex{r})$ to be real-valued.

In the clean limit, hybridizing the two surface Dirac cones to the flat band opens up (indirect)
spectral gaps, similar to the $x$-cut lattice model, see Fig.~\ref{fig:lattice_flat_spec}(b) for the hybridized spectrum of the latter.
The hybridized bands of the 2D continuum model are shown in Fig.~\ref{fig:ci_continu_spec}.

\subsubsection{Disorder}

Disorder is incorporated into the continuum model using the random-phase parameterization 
in momentum space \cite{Chou14,Ghorashi18},
\bsub \label{eq:ci_continu_a}
\begin{align}\label{NABVecDirt}
    A^i_{x, y} (\tmmathbf{q}) 
    =&\,
    \frac{\sqrt{\lambda_a}}{L} 
    \,
    \exp\left[{\mathi \, \Theta^i_{x, y} (\tmmathbf{q}) - \frac{q^2 \xi^2}{4}}\right],
\\
    A_{c} (\tmmathbf{q}) 
    =&\,
    \frac{\sqrt{\lambda_c}}{L} 
    \,
    \exp\left[{\mathi \, \Theta_{c} (\tmmathbf{q}) - \frac{q^2 \xi^2}{4}}\right],
\end{align}
\begin{align}
    \gamma_{1, 2} (\tmmathbf{q}) 
    =&\,
    \frac{\sqrt{\lambda_g}}{L} 
    \,
    \exp\left[\mathi \, \Theta_{1, 2}(\tmmathbf{q}) - \frac{q^2 \xi^2}{4}\right],
\end{align}
\esub
where $\vex{q}$ denotes a 2D momentum vector
and 
where each phase angle satisfies $\Theta(\vex{q}) = - \Theta(-\vex{q})$, 
but these are otherwise independent, uniformly distributed random variables over $[0,2\pi)$.
The disorder strengths are given by the variances 
\begin{itemize}
\item{$\lambda_a$: disorder that scatters between the two colors of the surface theory,}
\item{$\lambda_c$: disorder in the flat-band energy,}
\item{$\lambda_g$: disorder in the coupling between the surface theory and the 2D flat band.}
\end{itemize}
In Eq.~(\ref{eq:ci_continu_a}),
$L$ denotes the linear system size and $\xi$ is a correlation length of order the inverse momentum cutoff
\cite{Chou14}.

\subsection{Chiral Bistritzer-MacDonald model for TBLG \label{sec:cTBLG}}

Finally, we also consider the effects of chiral (``twist'' 
\cite{wilson2020disorder,Padhi20,shavit2023strain,nakatsuji2022moire,guerrero2025disorderinduced,sanjuanciepielewski2025transport,Queiroz25,
uri2020mapping,kapfer2023programming}) disorder on the chiral
limit of the BM model for TBLG. 
The clean continuum Hamiltonian is  \cite{Santos07,BM11,Tarnopolosky19} 
\begin{align}\label{TBLG}
     h_{\mathsf{BM}}^\pupsf{2D}  
     =&\, 
     - 
     \mathi 
     \,
     \vex{\sigma} \cdot \vex{\nabla} 
\nonumber\\
    &\,+ 
     \alpha 
     \left[
        U_1^{\ast}(-\vex{r}) \, \sigma^+ \, \tau^+  
        + 
        U_1(\tmmathbf{r}) \, \sigma^- \, \tau^+
        +
        \mathrm{H.c.}
    \right]\!,\!\!
\end{align}
where $U_1(\tmmathbf{r})$ is the periodic moir\'e potential
\begin{gather}\label{moirepot}
  U_1 (\tmmathbf{r}) 
  = 
  e^{- \mathi \tmmathbf{q}_1 \cdot \tmmathbf{r}} 
  +
  e^{\mathi \frac{2 \pi}{3}}
  e^{- \mathi \tmmathbf{q}_2 \cdot \tmmathbf{r}} 
  + 
  e^{- \mathi \frac{2 \pi}{3}} 
  e^{- \mathi \tmmathbf{q}_3 \cdot \tmmathbf{r}},
\NL 
  \tmmathbf{q}_1 = \left(0, -1\right), 
  \quad
  \tmmathbf{q}_2 = {\textstyle{\left(\frac{\sqrt{3}}{2}, \frac{1}{2}\right)}}, 
  \quad
  \tmmathbf{q}_3 = {\textstyle{\left(\frac{- \sqrt{3}}{2}, \frac{1}{2}\right)}}.
\end{gather}
Here $\{\vex{q}_{1,2,3}\}$ are the moir\'e reciprocal lattice vectors
(for the symmorphic moir\'e BZ) and $\alpha$ is the effective dimensionless 
interlayer tunneling, which gets larger for \emph{smaller} twist angles $\theta \ll \pi/2$ ($\alpha \sim 1/ \theta$ \cite{BM11}). 
The chiral BM model takes exactly the same form as the continuum 2-color topological surface
theory in Eq.~(\ref{hx2D}), with the periodic intercolor coupling
involving vector potential components $A_{x,y}^{1,2}$; the components
$A_{x,y}^3 = 0$. Therefore the chiral BM model is also in class CI,
satisfying 
\begin{align}
    - M_S \,  h_{\mathsf{BM}}^\pupsf{2D} \, M_S
    =
    - M_P \, \left(h_{\mathsf{BM}}^\pupsf{2D}\right)^\T \, M_P
    =  
    h_{\mathsf{BM}}^\pupsf{2D},
\end{align}
with $M_S = \sigma_3$ and $M_P = \sigma_1 \, \tau_2$. 

Perfect flat bands occur at zero energy at the first magic angle $\alpha \simeq 0.586$ \cite{Tarnopolosky19}. 
We incorporate uncorrelated disorder into each of the components of the periodic potential,
\begin{align}\label{TwistDirt}
  \delta h_{\mathsf{BM}}^\pupsf{2D}
  =&\,
  A_1 (\tmmathbf{r}) \, \sigma_1 \, \tau_1  
  + 
  A_2 (\tmmathbf{r}) \, \sigma_2 \, \tau_1  
\nonumber\\
  &\,
  +
  A_3 (\tmmathbf{r}) \, \sigma_1 \, \tau_2 
  + 
  A_4 (\tmmathbf{r}) \, \sigma_2 \, \tau_2.
\end{align}
These are the same as the $A_{x,y}^{1,2}$ disorder components in the continuum
surface model Eq.~(\ref{NABVecDirt}), and the disorder is implemented numerically in the same way. 
Eq.~(\ref{TwistDirt}) can be regarded as white noise ``twist disorder,'' because it involves
the same interlayer tunneling matrices as the BM model itself. We neglect the additional
effects of lattice relaxation, which would enter as an additional abelian vector potential
\cite{NetoGeim2009,shavit2023strain}
(and change the symmetry class to AIII).

\begin{figure}[t!]
  \centering
  \includegraphics[width=0.25\textwidth]{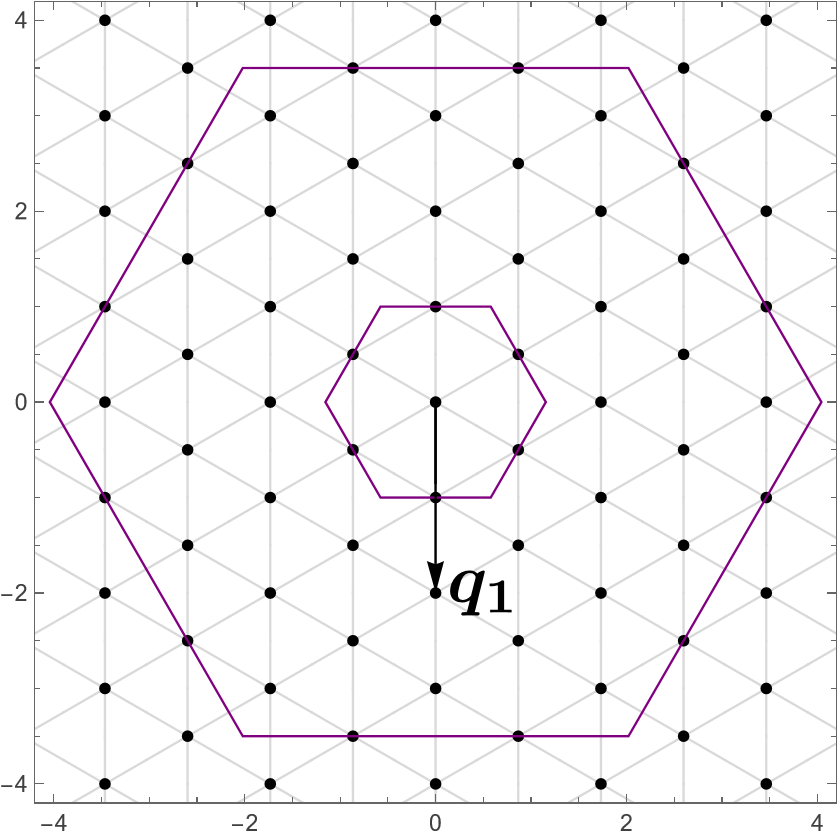}
  \caption{\label{fig:BM_Lattice}Schematic diagram of the momentum lattice used for the BM model
  calculation. The inner hexagon is the Brillouin zone of the moir\'e potential and the outer hexagon is the
  boundary of the momentum lattice.}
\end{figure}

In our subsequent numerical calculations for the BM model, we employ a triangular momentum-space lattice, which is shown in
Fig.~\ref{fig:BM_Lattice}. The boundary hexagon of the lattice has edge length $\frac{1}{\sqrt{3}}  \left(
\frac{5}{2} + M_1 \right)$. The lattice constant is ${1}/{(2 M_2)}$. The system size is $L^2$ with $L = 5
M_2 + 2 M_1 M_2$.

\section{Fragility to localization} \label{sec:result}

\subsection{Multifractal measures}

To characterize the degree of Anderson localization for surface-state wave functions, we
employ multifractal analysis \cite{Evers08}. By summing moments of the position-space probability for eigenstates over the surface, we obtain scaling behavior with respect to the surface linear system size $L$. Anderson localized states with localization length smaller than the system size give moments that do not scale with $L$, while extended states scale as a simple power of $1/L$. Critical wave functions (such as the topologically protected zero-energy states
\cite{Mudry1996,Caux1996,Schnyder08,Essin15,Schulz-BaldesBook}
as well as finite-energy surface states in the SWQC scenario \cite{Ghorashi18,Sbierski20,ghorashi2020criticality,Karcher21,UFO24,Zhang25}) 
exhibit nontrivial power-law scaling with $1/L$, encoded in the so-called multifractal spectrum of scaling exponents. We introduce these in the context of surface states obtained for the bulk 3D lattice model. 

For the lattice model with the $z$-cut [Fig.~\ref{fig:ci_surf_spec}(a)], we define the probability at site position $(x,y)$ on the surface for a surface eigenstate wave function $\psi_\e(x,y,z)$ 
by summing the probability through a depth of $n_1$ layers (and renormalizing),
\begin{align}\label{SurfaceProb}
    p(x,y;\e) 
    \equiv
    \frac{
    \sum_{z = 1}^{n_l} 
    |\psi_\e(x,y,z)|^2    
    }{
    \sum_{z' = 1}^{n_l} \sum_{x',y'} |\psi_\e(x',y',z')|^2
    }.
\end{align}
We then compute the box probability
\begin{equation}
    \mu_b (x, y;\e) 
    = 
    \sum_{(x', y') \in b(x, y)} p (x', y';\e),
\end{equation}
where $b(x,y)$ denotes a square of linear size $b$ centered at $(x,y)$. 
This coarse-graining step allows us to extract multifractal dimensions
from scaling with box sizes in a fixed system size $L$.
These are determined from the inverse participation ratio
\begin{equation}\label{IPR}
    P_b (q;\e) 
    = 
    \sum_{x, y} 
    \left[\mu_b(x, y)\right]^q 
    \propto 
    \left( \frac{b}{L}\right)^{\tau (q)},
\end{equation}
where $\tau(q)$ is the multifractal exponent. 
Unlike $P_b(q;\e)$, for weak multifractality the exponent $\tau(q)$ is self-averaging  
in the infinite-$L$ limit. 
At a certain $q$, we can fit $\ln P_b (q;\e)$ averaged over $N_s$ 
energy-adjacent eigenstates
with $\ln (b / L)$ under several different box sizes $\{b\}$ to extract the anomalous part
of the spectrum,
\begin{align}
    \Delta (q) = \tau (q) - 2 (q - 1).
\end{align}
For weak multifractality and small $q$, this can be approximated by \cite{Chou14}
\begin{equation}\label{thetaDef}
  \Delta(q) = - \theta(\e) \, q (q - 1).
\end{equation}
From the fitted $\tau(q)$, we can extract $\theta(\e)$ near energy $\e$ 
[and determine the validity of the parabolic approximation to the spectrum $\tau(q)$]. 
When the states are extended and ergodic, $\theta=0$. 
When the states are critical, $\theta > 0$ is finite and (approximately) $q$ independent. 
Anderson localized states have $\tau(q > 0) = 0$, which formally corresponds to 
$\theta = \theta_q \rightarrow 2/q$.

\subsection{Surface-state scenarios}

As discussed in the Introduction, localizable topological phases such as class CI in 3D \cite{UFO24}
have fragile surface states that can be destroyed (e.g., Anderson localized by \emph{weak} disorder). 
The only exceptions are the protected zero-energy surface states, but these form a set of measure zero in the surface-state spectrum.
The localization of all but the zero-energy surface states has important implications for experimental observables;
for example, such an almost-localized surface cannot produce a finite surface current at finite temperature 
(in the absence of interaction-mediated inelastic scattering) \cite{Wang2000}.

In fact, previous numerical studies of class-CI continuum Dirac surface states instead
demonstrated robust spectrum-wide quantum criticality (SWQC) \cite{Ghorashi18,Karcher21}, 
reviewed in Secs.~\ref{sec:SWQC} and \ref{sec:resultSum}, see also Fig.~\ref{fig:theta}. 
In particular, it was observed that finite-energy surface states of the Dirac
model in Eq.~(\ref{hx2D}) exhibit \emph{universal} fractality, with $\theta \simeq 1/8$,
consistent with the known value of the class-C \emph{spin} quantum Hall plateau transition (SQHPT).
This is not to be confused with the quantum spin Hall effect and 2D $\mathbb{Z}_2$ topological insulators
\cite{BernevigBook}; the spin quantum Hall effect is an analog of the ordinary integer effect 
predicted to occur in a topological 2D $d + i d$ superconductor \cite{QiZhang2011,Evers08}. The plateau transition
then involves bulk quasiparticle states that carry spin, instead of charge (the latter
being ill-defined for quasiparticles in a superconducting state). 
The observation of class-C states is in and of itself a persistent mystery, at odds with standard
arguments that finite-energy states of 10-fold way Hamiltonians always reside in one of the three Wigner-Dyson classes A, AI, or AII \cite{Evers08}.
A symmetry argument in favor of class C for finite-energy \emph{Andreev bound} surface states of a class-CI topological phase,
based upon replica-sigma-model target manifolds, was advanced in Ref.~\cite{Ghorashi18}.
We will not dwell on this point here, but return to it as an open question in the conclusion Sec.~\ref{sec:oq}.

Continuum Dirac surface states of the localizable class AIII were likewise found to exhibit 
ordinary integer quantum Hall criticality \cite{Sbierski20}. In Ref.~\cite{UFO24},
it was shown however that this requires statistical fine-tuning in a microscopic
bulk lattice model of a class AIII topological phase. In particular, a certain ``fragmenting potential''
was identified in Ref.~\cite{UFO24} with unexpected properties.
The fragmenting potential
can be turned on as a uniform perturbation at the surface of the lattice
model that preserves the AIII symmetry,
and induces the following effects:
(1) it can open a gap between the surface and bulk spectra,
(2) it pushes Berry curvature into the lattice-model surface states,
(3) it enables Anderson localization of almost all surface states (by chiral symmetric disorder).
And yet, this mechanism is completely absent in the continuum Dirac models for AIII surfaces, because
the Berry curvature induced by the fragmenting potential goes into the lattice ``microstructure''
of the surface states, neglected in continuum treatments \cite{UFO24}.

\begin{figure}[b!]
  \centering
  \subfigure[]{
    \includegraphics[width=0.4\textwidth]{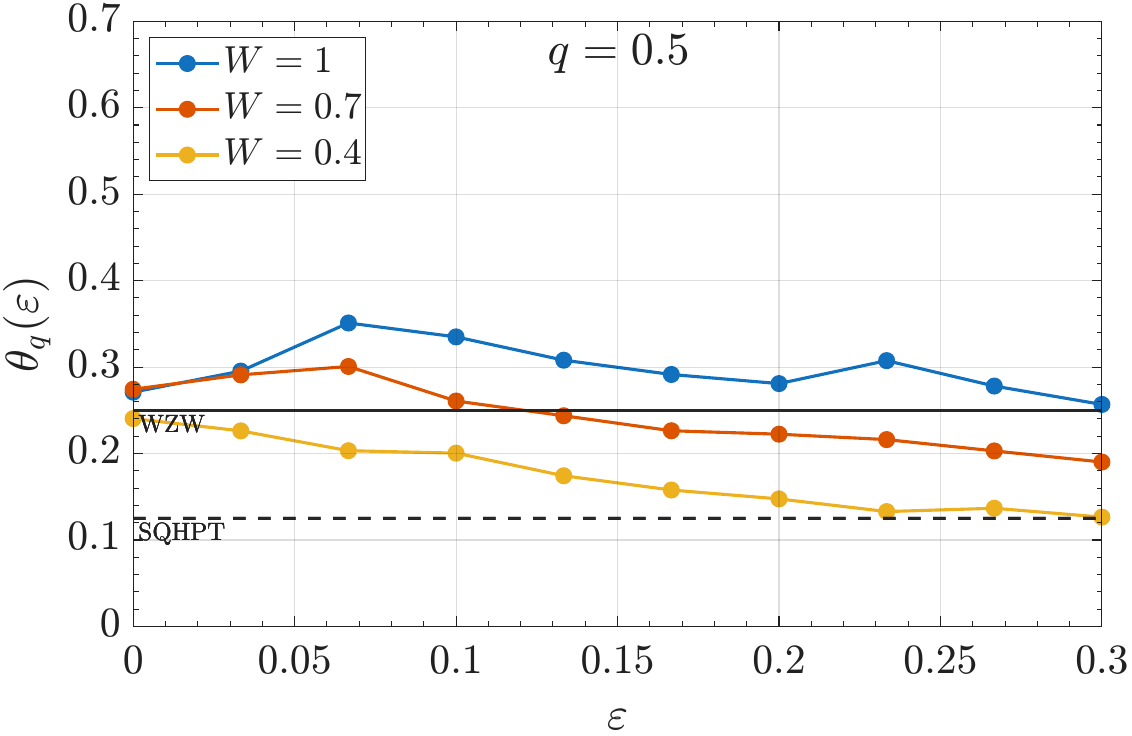}
  }
  \subfigure[]{
    \includegraphics[width=0.4\textwidth]{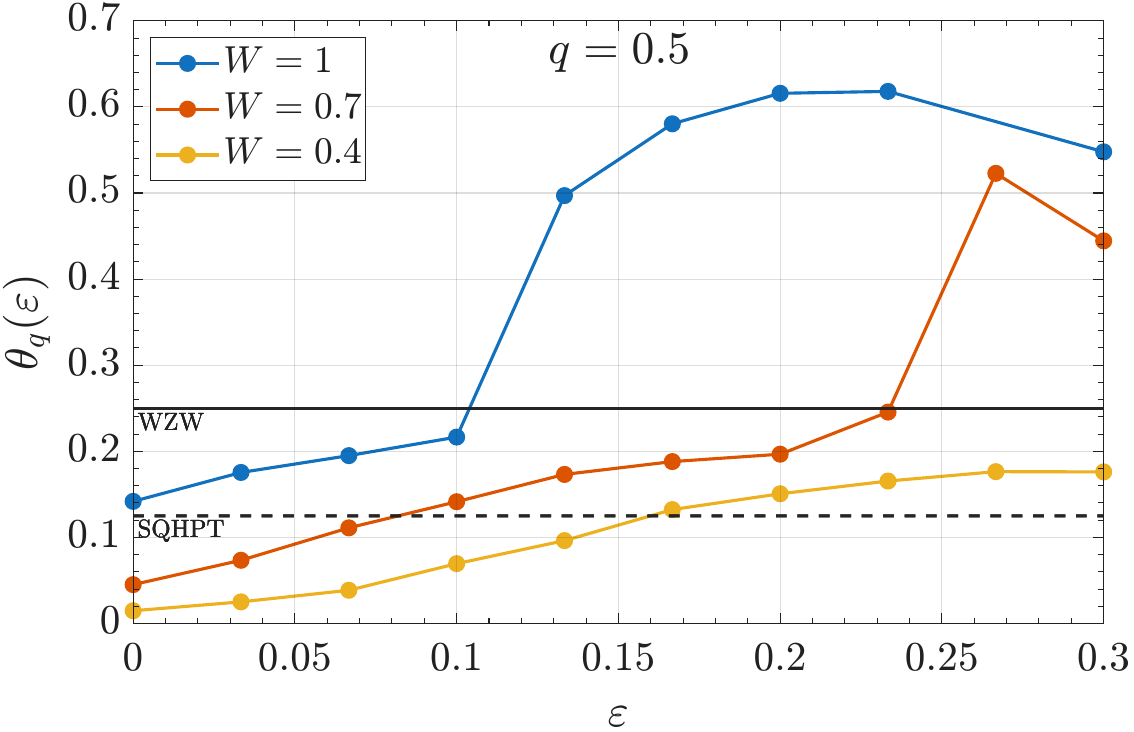}
  }
  \caption{\label{fig:ci_lattice_theta_dis}
  Effects of disorder on the surface states of the CI lattice model 
  [Eqs.~(\ref{eq:hamiltonian}) and (\ref{eq:HamTPE})]
  and the topological proximity effect (TPE).
  We plot $\theta_q (\varepsilon)$ 
  [Eq.~(\ref{thetaDef})]
  without and with coupling to the flat band 
  for varying disorder strengths.  
  (a) Surface states without coupling to the flat band. 
  The weak effect of increasing the disorder strength on $\theta_q(\e)$ 
  as well as 
  its approximate energy independence
  are qualitatively consistent with the SWQC scenario \cite{Ghorashi18,Karcher21}.
  (b) Demonstration of the TPE: lattice surface states hybridized with the trivial flat band. 
  The trend of increasing $\theta_q(\e)$ with increasing disorder for the finite-energy states 
  indicates a drift towards Anderson localization for these states. 
  The overall disorder strength is $W$ (see text) for each curve. 
  Other parameters are $N_y = 72$, $N_z = 10$, $\mu = - 2$, $\Delta_1 = 1$, $\Delta_2 = 1$, $m_c = 1$,
  and for panel (b)  
  $\varepsilon_c = 0.4$ and $\gamma = 0.5$.
  }
\end{figure}

We focus here on an alternative way to destroy surface states proposed in Ref.~\cite{UFO24},
which is coupling the surface to trivial degrees of freedom in a way that preserves the defining
symmetry of the class (trivializing proximity effect, TPE). This mechanism can be employed in both bulk lattice and 2D continuum Dirac models,
and enables the opening of a gap in the surface-state spectrum of both. 
Our goal is to determine under what circumstances (if any) the SWQC scenario persists in the presence
of TPE.

\begin{figure}[b!]
  \centering
  \subfigure[]{
    \includegraphics[width=0.4\textwidth]{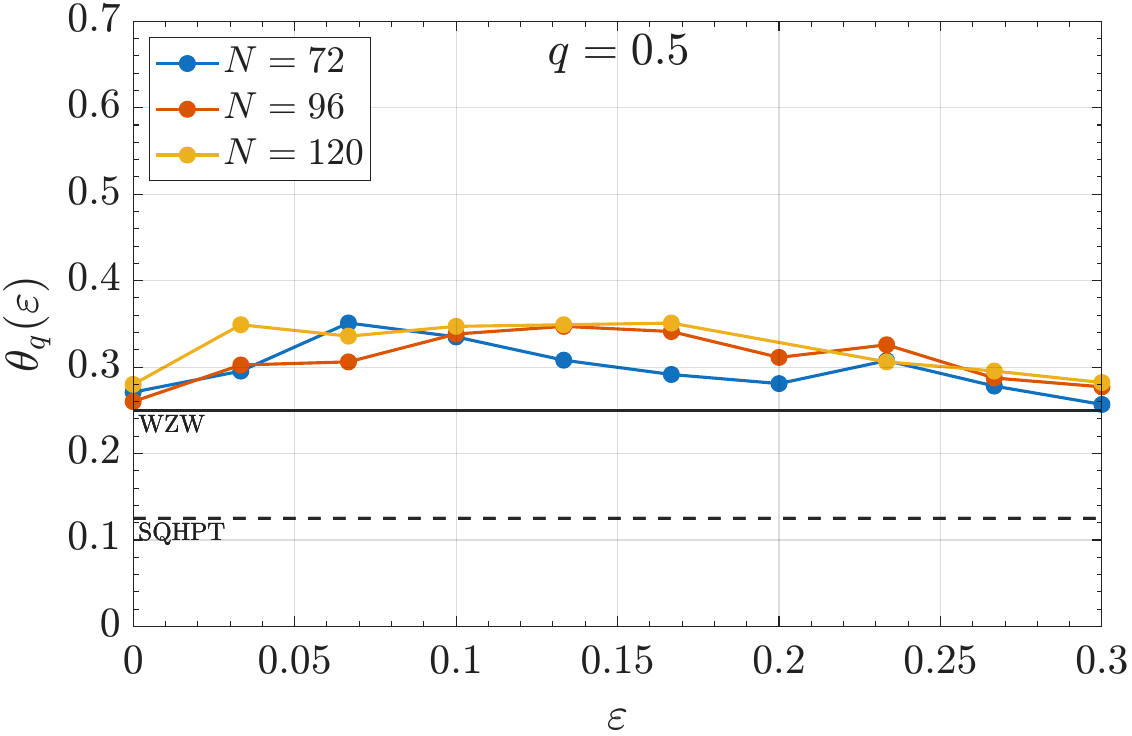}
  }
  \subfigure[]{
    \includegraphics[width=0.4\textwidth]{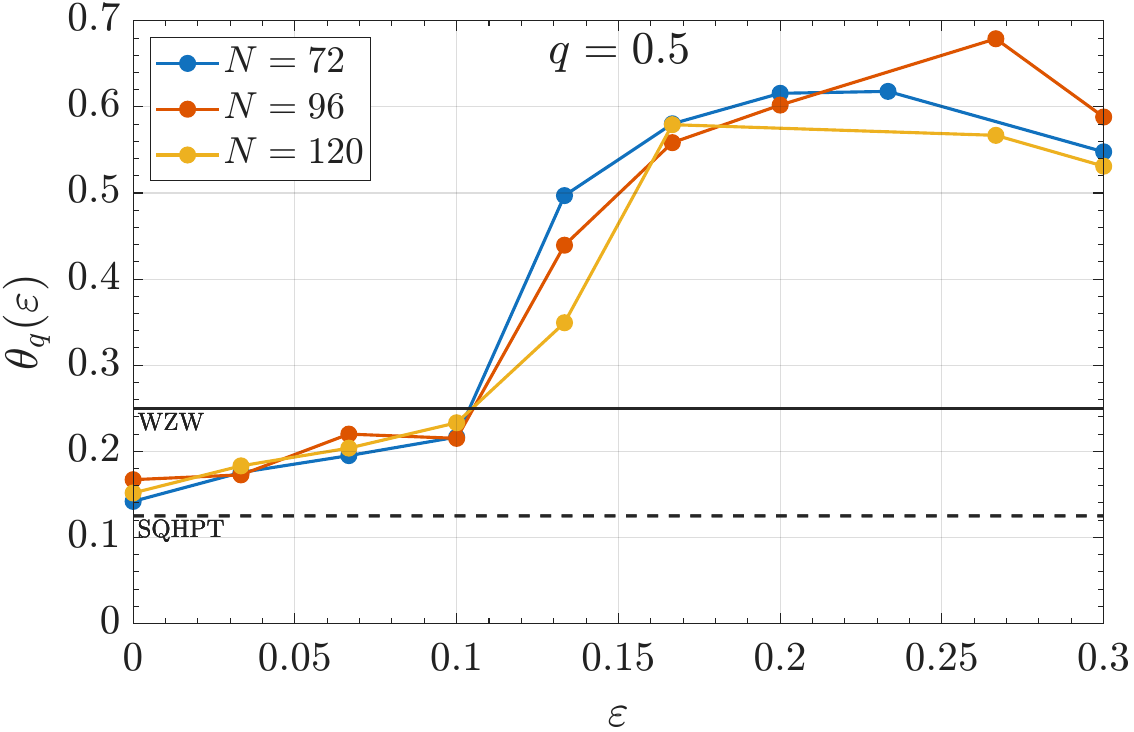}
  }
  \caption{\label{fig:ci_lattice_theta_n} 
  The same as Fig.~\ref{fig:ci_lattice_theta_dis}, but for fixed disorder and three different system sizes. 
  (a) Surface states in the absence of the flat band. 
  (b) Surface states hybridized with the flat band. 
  The parameters are $W = 1$, 
  $N_z = 10$, $\mu = - 2$, $\Delta_1 = 1$,
  $\Delta_2 = 1$, $m_c = 1$,
  and for panel (b)
    $\varepsilon_c = 0.4$, $\gamma = 0.5$.  
  }
\end{figure}

\subsection{CI cubic lattice model, TPE}

We examine the surface states of the lattice model in Eq.~(\ref{eq:hamiltonian}), without and with coupling
to the trivial flat band encoded in Eq.~(\ref{eq:HamTPE}). We calculate the surface states of an 
$N_x$$\times$$N_y$$\times$$N_z$ thin slab, with $N_x = N_y \gg N_z$. 
In order to focus on the low-energy surface states of the top $z$-cut interface, 
we gap out the bottom one with a surface Chern mass $m_c \, \tau_2$.
This breaks the CI symmetry down to C on the bottom layer, but has negligible effect on the bulk or top surface
for the thicknesses $N_z$ employed here. 

Since we study the $z$-cut surface, the clean system without the flat-band coupling has the quadratically
dispersing surface states (ala Bernal-stacked bilayer graphene) shown in Fig.~\ref{fig:ci_surf_spec}(a). 
Unlike class AIII, we were unable to identify a uniform surface modification 
that does not involve additional degrees of freedom, 
preserves the class-CI symmetry, 
and which opens up a gap between surface and bulk spectra. In other words
there seems to be no homogeneous analog of the fragmenting potential employed in Ref.~\cite{UFO24}. 
An analogous term might exist for the $x$-cut surfaces, but likely requires spatial modulation 
at the wave vector connecting the two surface Dirac cones shown in Fig.~\ref{fig:ci_surf_spec}(b);
again we have not found a homogeneous potential perturbation that can ``detach'' the surface states. 

We calculate the multifractal curvature parameter $\theta$ [Eqs.~(\ref{IPR})--(\ref{thetaDef})]
for surface states subject to disorder, with and without the flat band. We present results for
varying disorder strengths and systems sizes. 
Disorder with uniform distribution $[-W, W]$ is added to each of the parameters $\mu$, $\Delta_1$,
$\Delta_2$, $\varepsilon_c$, and $\gamma$ in Eq.~(\ref{eq:hamiltonian}) and Eq.~(\ref{eq:HamTPE}). The
ratio between the disorder strengths is kept constant while the overall disorder magnitude is varied.
Results are exhibited in Figs.~\ref{fig:ci_lattice_theta_dis} and \ref{fig:ci_lattice_theta_n}.
Here the depth used to define the surface states is $n_l = 1$ without the flat band, 
and $n_l = 3$ with it (including the flat band layer) [Eq.~(\ref{SurfaceProb})]. 
Each $\theta_q(\e)$ at an energy point $\e$ is an average from 30 states immediately around that energy. The wave
functions are computed using kernel polynomials 
and the filter and shake method \cite{weisse2006kernel, guan2021dual,Zhang25}. 

Disorder is incorporated into both the flat band and top layer of the crystal in order to minimize finite-size effects,
such as the presence of ballistic (plane-wave-like) states. For an effectively 2D disordered class-CI system, only quantum-critical or Anderson-localized wave functions are possible in the thermodynamic limit \cite{Evers08}. We diagnose the appropriate strength of disorder by examining the criticality of the $E = 0$ surface state, a strategy that has worked well in the past \cite{Ghorashi18}. We are also careful to avoid making the disorder too strong so as to prevent the trivial localization of the outer layer of the crystal; in that case, the ``true'' surface states are simply submerged to a deeper layer. Weaker disorder is included in the bulk to avoid a strong discontinuity in the impurity profile, but plays no other role here. 

Figs.~\ref{fig:ci_lattice_theta_dis}(a) and \ref{fig:ci_lattice_theta_n}(a) demonstrate that the surface states 
of the lattice model in the absence of the coupling to the flat band
are stable with increasing disorder or system size, with small nonzero values of $\theta$ indicating wave function criticality. This is qualitatively consistent with the SWQC scenario, although the obtained $\theta(\e)$ are generally closer to the zero-energy analytical result $\theta = 1/4$, instead of the class-C SQHPT value $\theta \sim 1/8$. 
Calculations for the 2D continuum Dirac model shown in Fig.~\ref{fig:ci_conti_theta} instead converge to the class-C value, consistent with previous work \cite{Ghorashi18}. Finite-size effects are a possible explanation for this lattice versus continuum quantitative discrepancy (without the flat band), given that the continuum calculations effectively simulate a much larger system for the same numerical surface grid size. 

By contrast, Figs.~\ref{fig:ci_lattice_theta_dis}(b) and \ref{fig:ci_lattice_theta_n}(b) demonstrate the TPE: 
the coupling to the flat band enables disorder to push finite-energy surface eigenstates towards Anderson localization. 
Fig.~\ref{fig:ci_lattice_theta_dis}(b) shows the effect increases strongly with increasing disorder. This is one of the main results of this paper, and is consistent with the arguments advanced in Ref.~\cite{UFO24}. Note that the lowest-energy surface states remain critically delocalized, as expected \cite{Schnyder08,Essin15,Schulz-BaldesBook}.
Fig.~\ref{fig:ci_lattice_theta_n}(b) on the other hand indicates only minimal changes to the observed TPE with increasing system size. 

The sharp increase of $\theta_q(\e)$ near $\e = 0.1$ in Fig.~\ref{fig:ci_lattice_theta_dis}(b) for $W = 1$ 
indicates a crossover from critical, zero-energy behavior to finite-energy localization; this occurs in finite size when the localization length $\zeta(\e)$ (which diverges as $\e \rightarrow 0$ \cite{Ludwig1994}) becomes of order the system size $L$. For the largest system size in Fig.~\ref{fig:ci_lattice_theta_n}(b), our data is quite sparse.
This is because the filter and shake KPM algorithm struggles to converge for localized states and big system sizes;
compare to Fig.~\ref{fig:ci_lattice_theta_n}(a) (without the flat band), which shows robust SWQC and converges without difficulty for all system sizes.

For class AIII, it was shown in Ref.~\cite{UFO24} that coupling the surface states to a trivial flat band pushes Berry curvature into those states, in a similar fashion as the fragmenting potential. Surface domains with different signs of the fragmenting potential can induce 1D chiral edge modes ``in the sky,'' i.e.\ in the induced gap between bulk and surface states. By contrast, we have not found the analog of a fragmenting potential, and chiral edge states cannot span the gap with the class-CI $T^2 = + 1$ time-reversal symmetry intact. Despite the tacit identification of SWQC wave functions in the continuum model of class CI surface states \cite{Ghorashi18} with the spin quantum Hall plateau transition, there does not appear to be a mechanism for inducing chiral edge modes on the CI surface without changing the symmetry class. 
Thus the exact mechanism for the TPE in this case must be different from class AIII.


\begin{figure}[b!]
  \centering
  \subfigure[]{
    \includegraphics[width=0.35\textwidth]{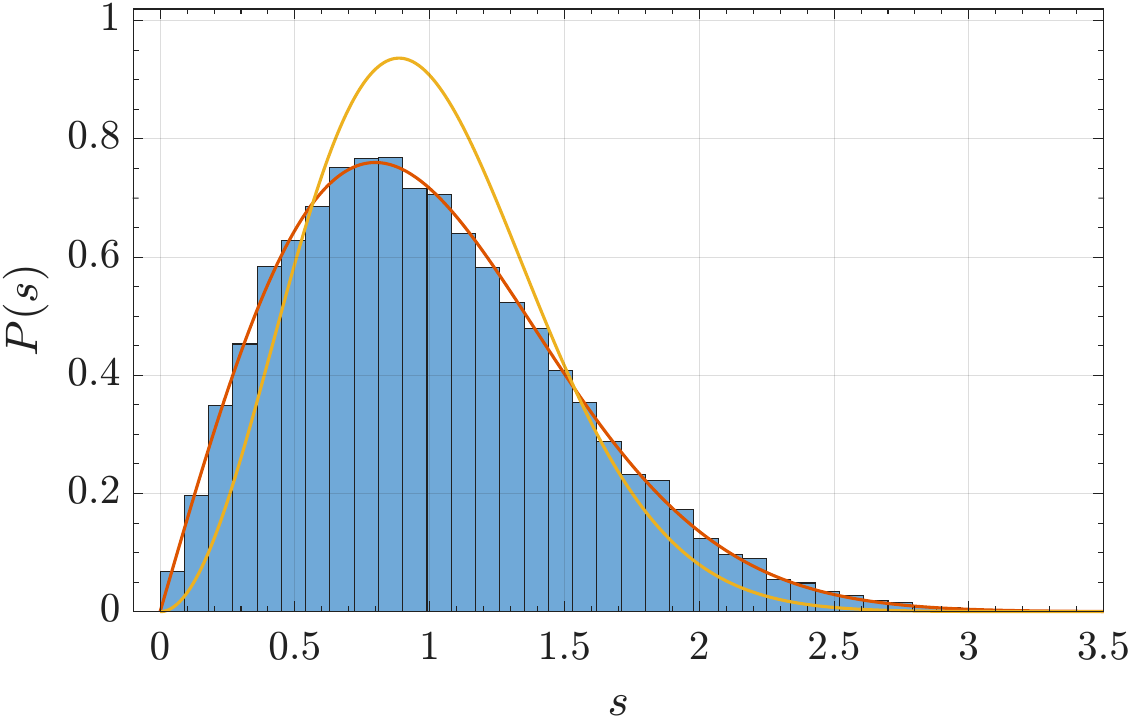}
  }
  \subfigure[]{
    \includegraphics[width=0.35\textwidth]{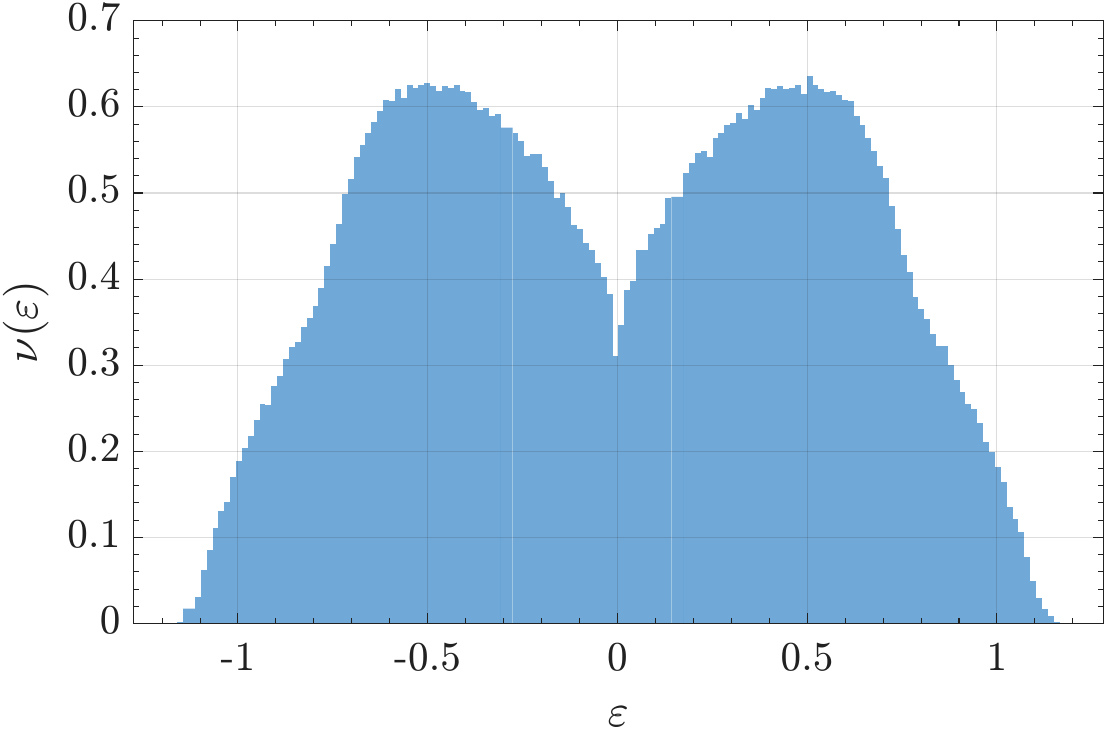}
  }
  \caption{\label{fig:ci_continu} Benchmark results for the CI continuum (2D) Dirac model 
  with disorder [Eq.~(\ref{hx2D})],
  without coupling to the flat band. 
  (a) Level statistics compared to the GOE (red) and GUE (orange) predictions. 
  The results are consistent with the GOE.
  (b) Global density of states $\nu(\e)$. The cusp near zero energy is expected \cite{Nersesyan1994}, 
  and was previously numerically analyzed and found to agree well with the analytical prediction
  \cite{Ghorashi18}.
  The linear system size is $L = 97$, 
  while the disorder strength and correlation length 
  [Eq.~(\ref{NABVecDirt})]
  are set to $\lambda_a = 2 \pi$ and $\xi = \pi$,
  respectively}
\end{figure}

\begin{figure}[t!]
  \centering
  \subfigure[]{
    \includegraphics[width=0.4\textwidth]{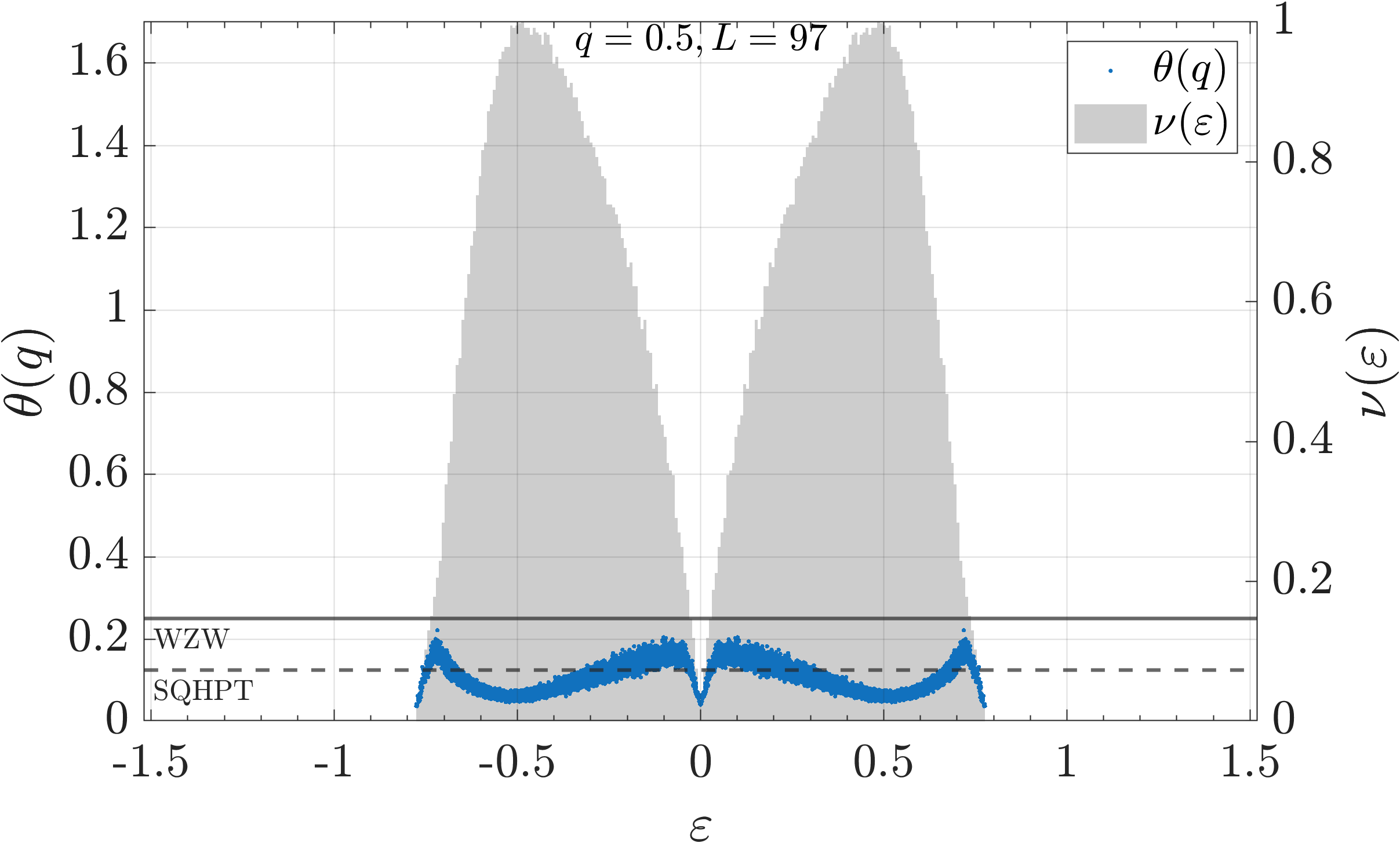}
  }
  \subfigure[]{
    \includegraphics[width=0.4\textwidth]{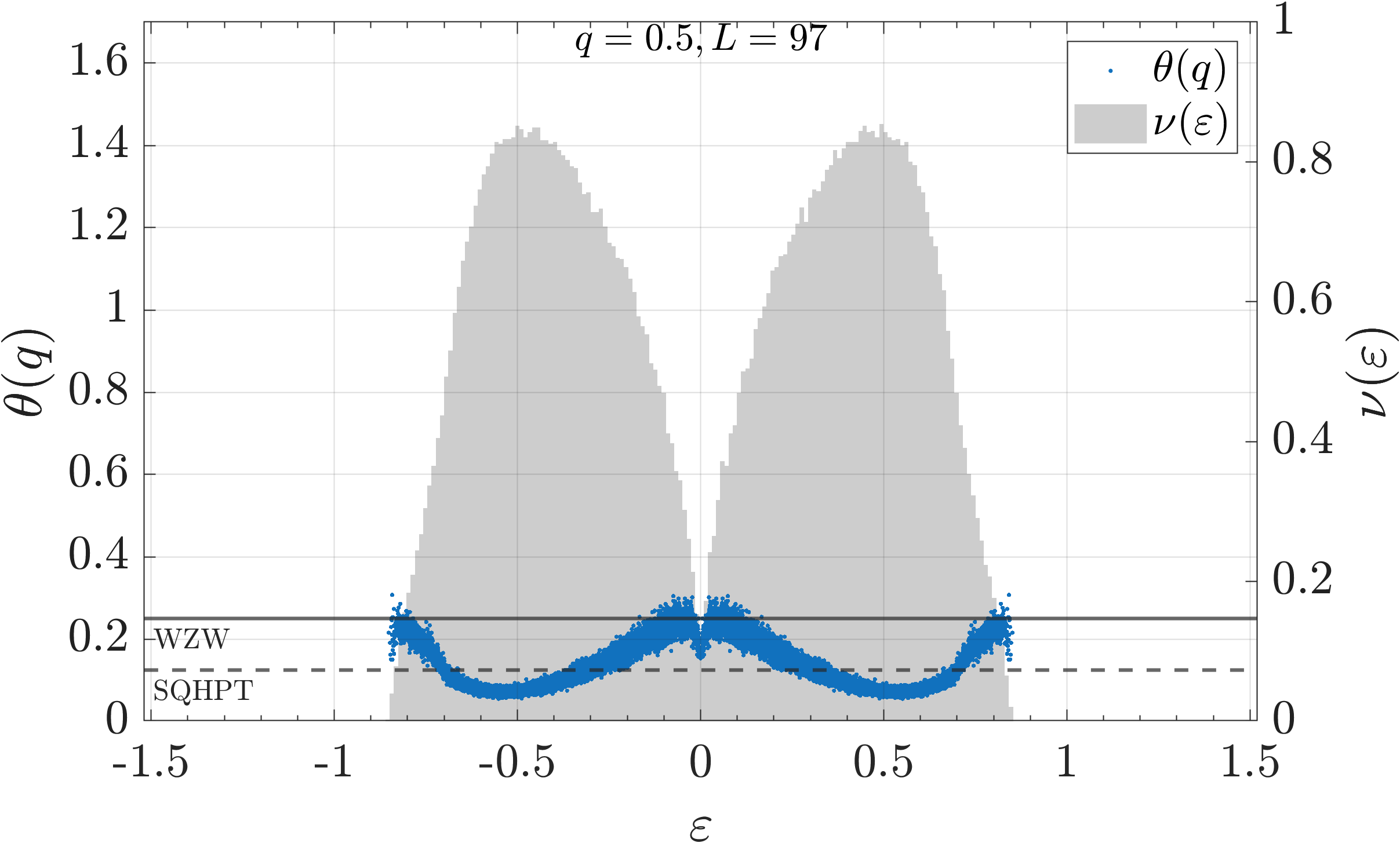}
  }
  \subfigure[]{
    \includegraphics[width=0.4\textwidth]{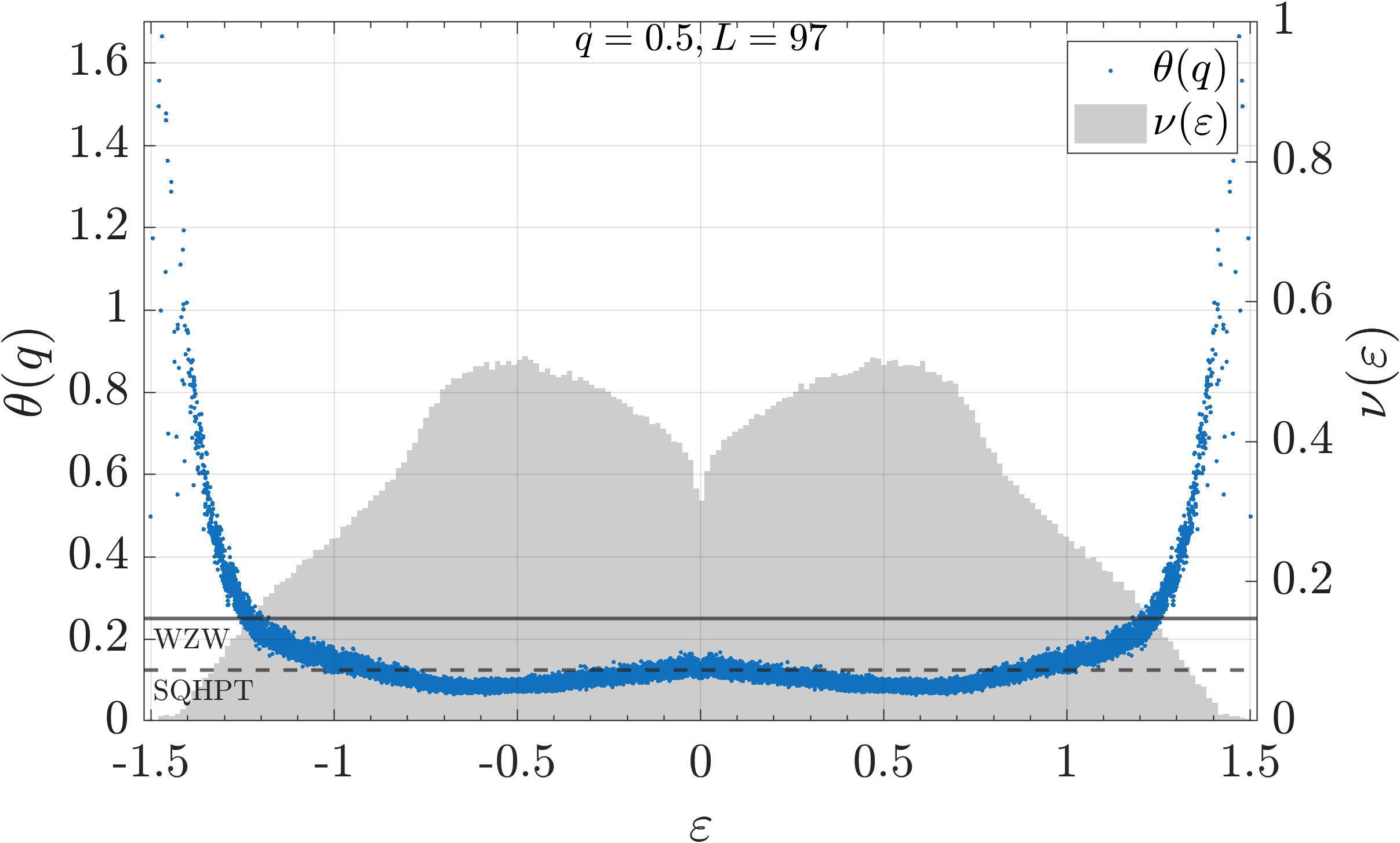}
  }
  \caption{\label{fig:ci_conti_theta} 
  Robust surface states of the  CI continuum Dirac model
  [Eq.~(\ref{hx2D})] without the flat band, demonstrating SWQC \cite{Ghorashi18,Karcher21}. 
    We plot the multifractal curvature $\theta_q (\varepsilon)$ [Eq.~(\ref{thetaDef})] and the corresponding global density
    of states $\nu(\e)$ versus eigenstate energy. 
    Results are shown for three different disorder strengths  [Eq.~(\ref{NABVecDirt})] in (a--c): $\lambda_a = 0.5$, 1, $2 \pi$. 
    Increasing the disorder strength creates more states with apparent class-C SQHPT critical statistics, as found
    previously in Ref.~\cite{Ghorashi18}.
    The other parameters are $L = 97$, $\xi = \pi$, and the boxes used for multifractal analysis are $b = (4, 6, 8)$.}
\end{figure}

\begin{figure}[t!]
  \centering
  \subfigure[]{
    \includegraphics[width=0.4\textwidth]{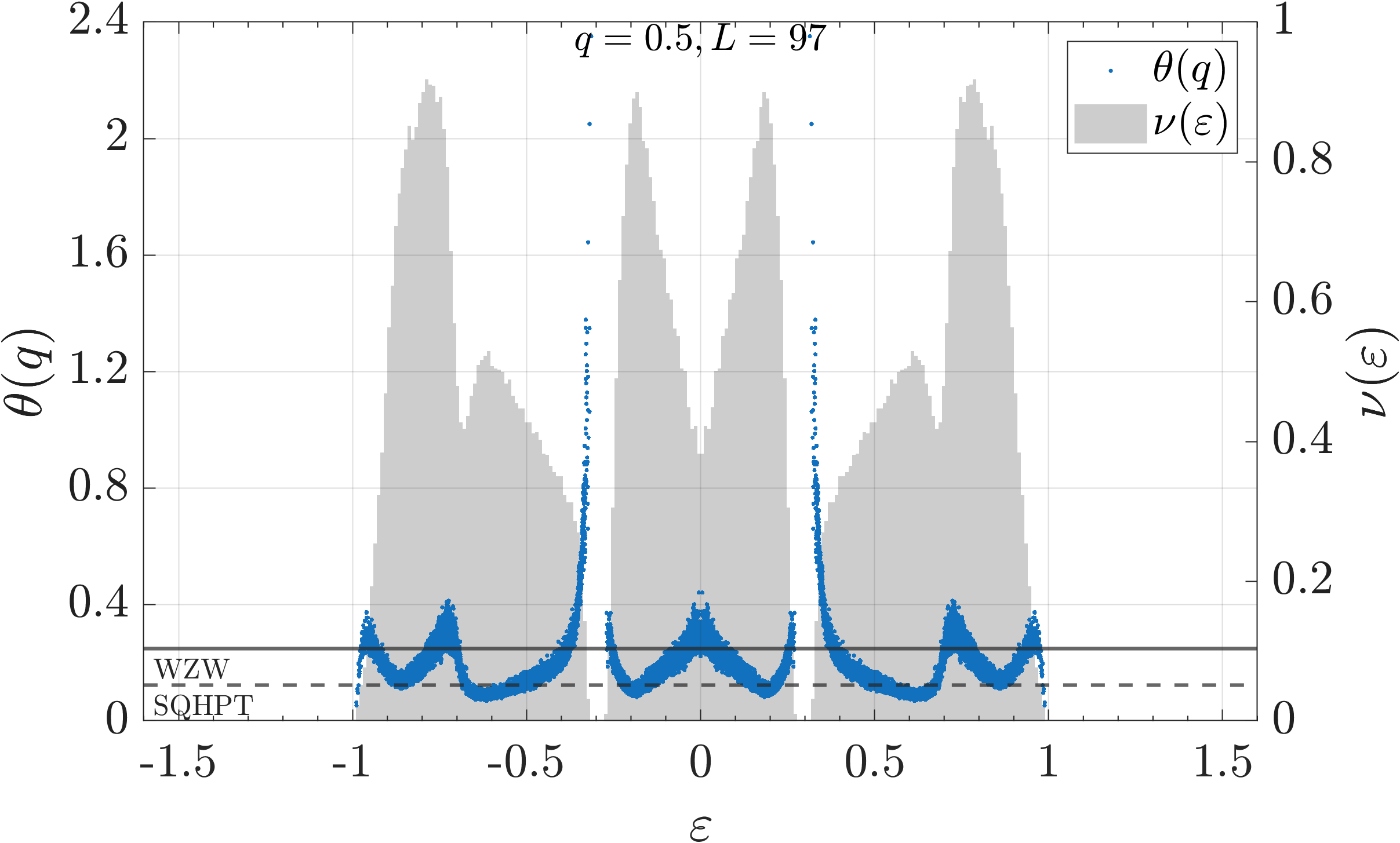}
  }
  \subfigure[]{
    \includegraphics[width=0.4\textwidth]{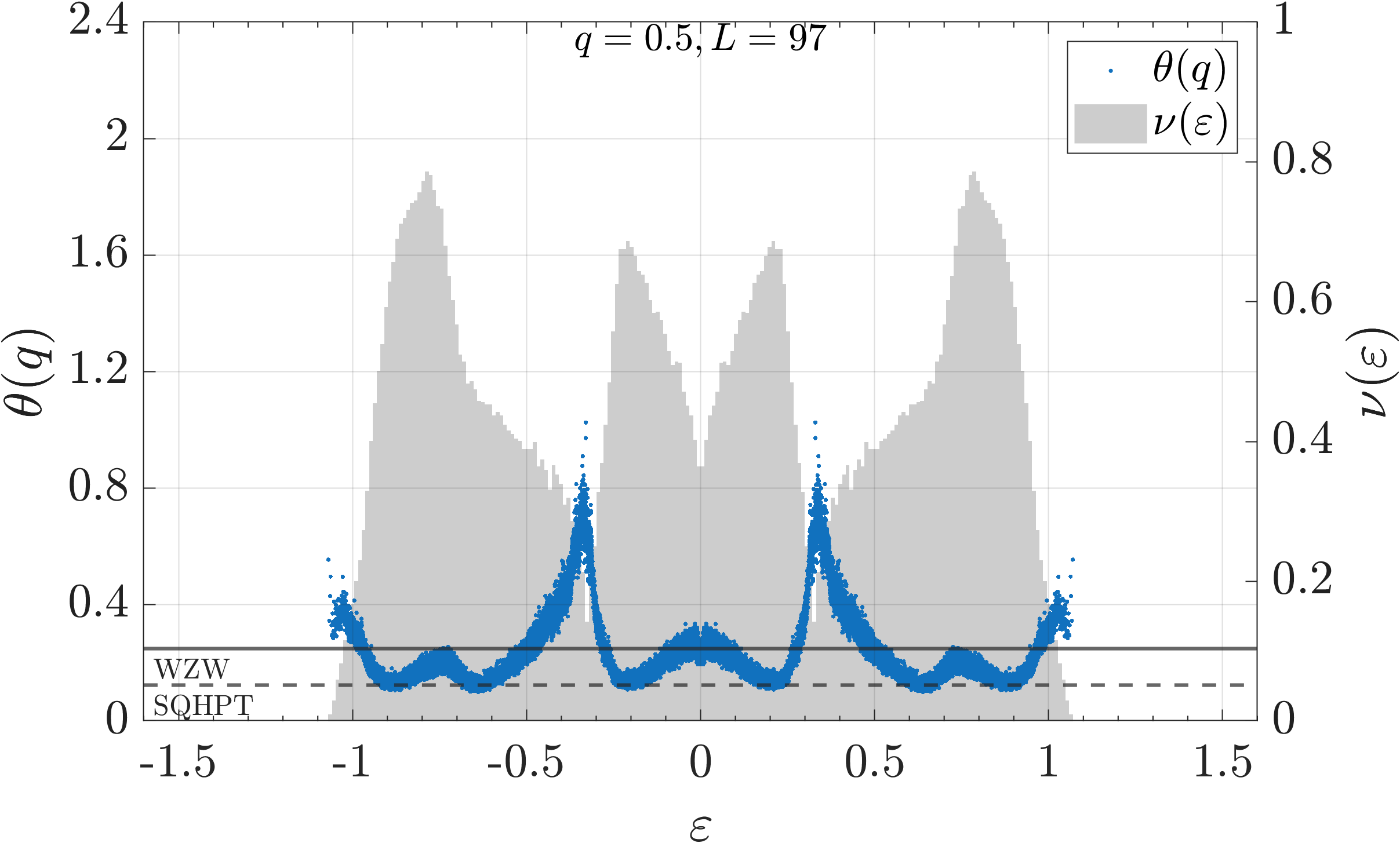}
  }
  \subfigure[]{
    \includegraphics[width=0.4\textwidth]{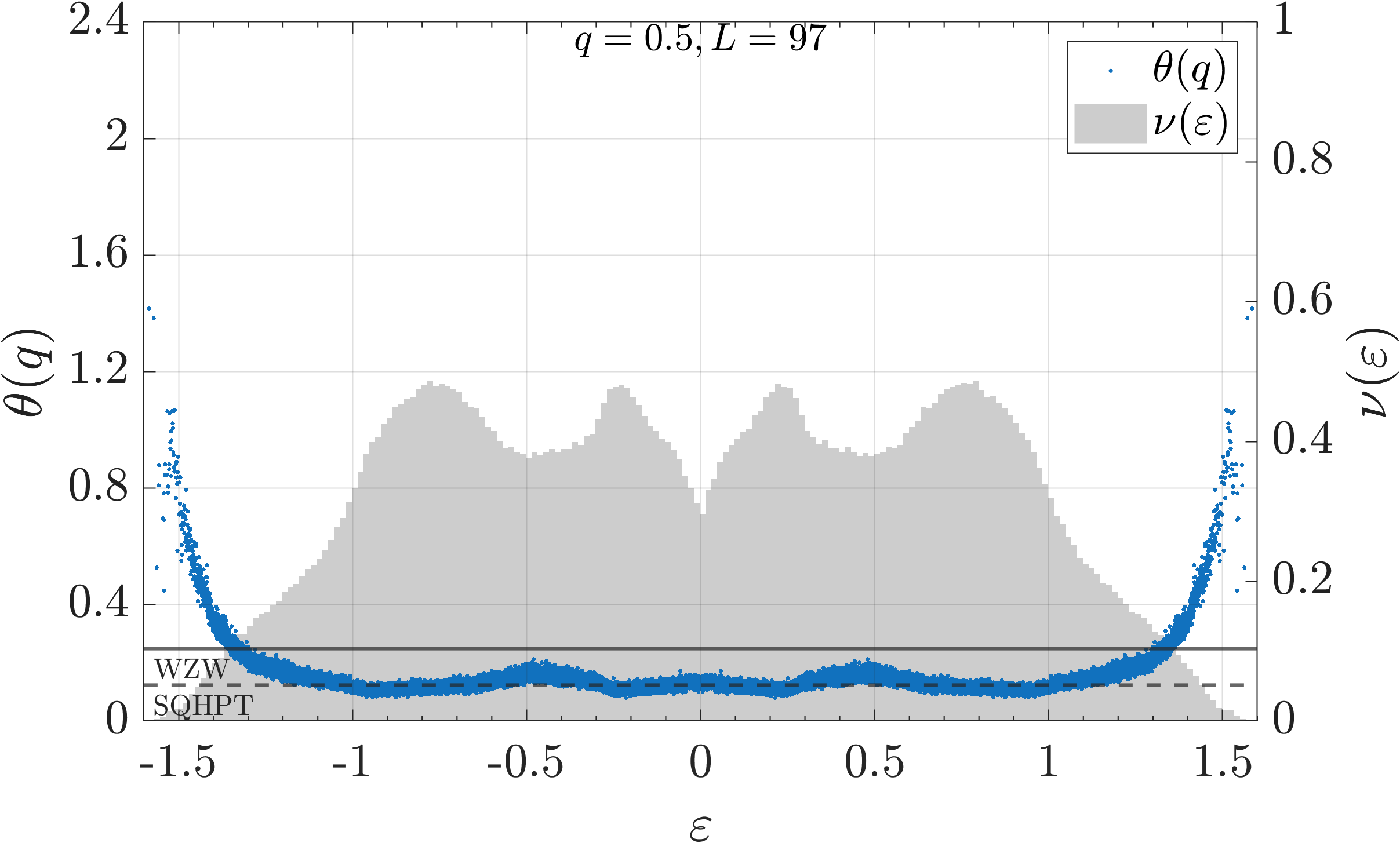}
  }
  \caption{\label{fig:ci_conti_flat_theta}
  Testing whether the continuum model exhibits the same TPE found for lattice surface states,
  see Figs.~\ref{fig:ci_lattice_theta_dis} and \ref{fig:ci_lattice_theta_n}.
    The class-CI 2D Dirac model is hybridized with the continuum flat band, Eq.~(\ref{HDiracTPE}).
    The coupling to the flat band opens up gaps in the clean surface spectrum,
    as shown in Fig.~\ref{fig:ci_continu_spec}.
    Here we incorporate disorder and examine the effects upon wave function fractality and the spectral gaps.
    For weak disorder $\lambda_a = 0.2$ (a), the spectral gaps persist and states near the gap edges
    are strongly localized with $\theta_q(\e) > 1$. 
    With increasing disorder $\lambda_a = 1$ (b) and $\lambda_a = 2 \pi$ (c),
    however, the spectral gaps wash out and SWQC appears to be restored---compare
    to Fig.~\ref{fig:ci_conti_theta}(c).
    These results indicate that the TPE is \emph{not} robust in the continuum model. 
  The other parameters are $\varepsilon_c = 0.4$, $\lambda_c = 0.2$, $\gamma_0 =
  0.35$, $\lambda_g = 0.2$, $L = 97$, $\xi = \pi$, $b = (4, 6, 8)$.}
\end{figure}


\begin{figure}[b!]
  \centering
  \subfigure[]{
    \includegraphics[width=0.4\textwidth]{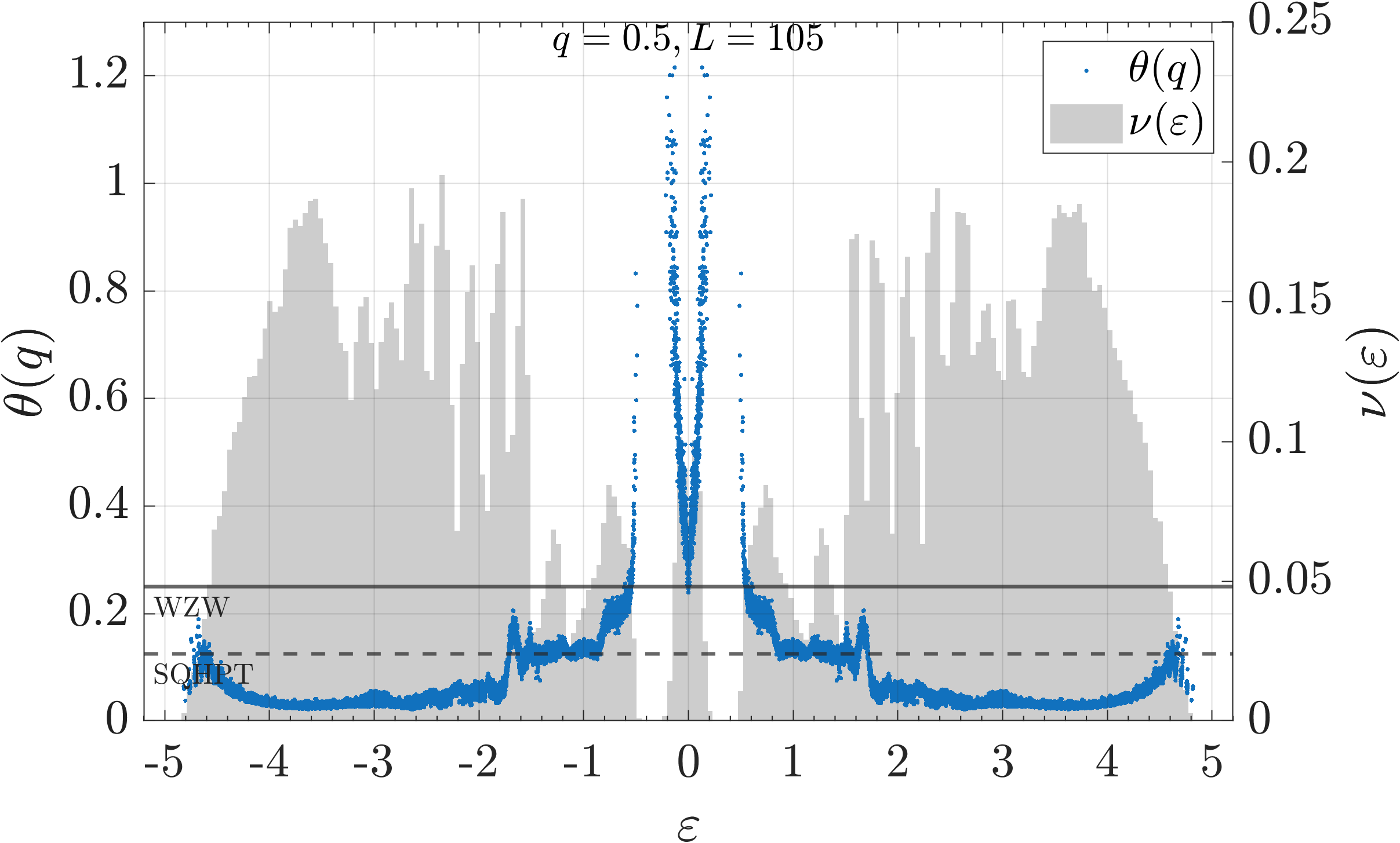}
  }
  \subfigure[]{
    \includegraphics[width=0.4\textwidth]{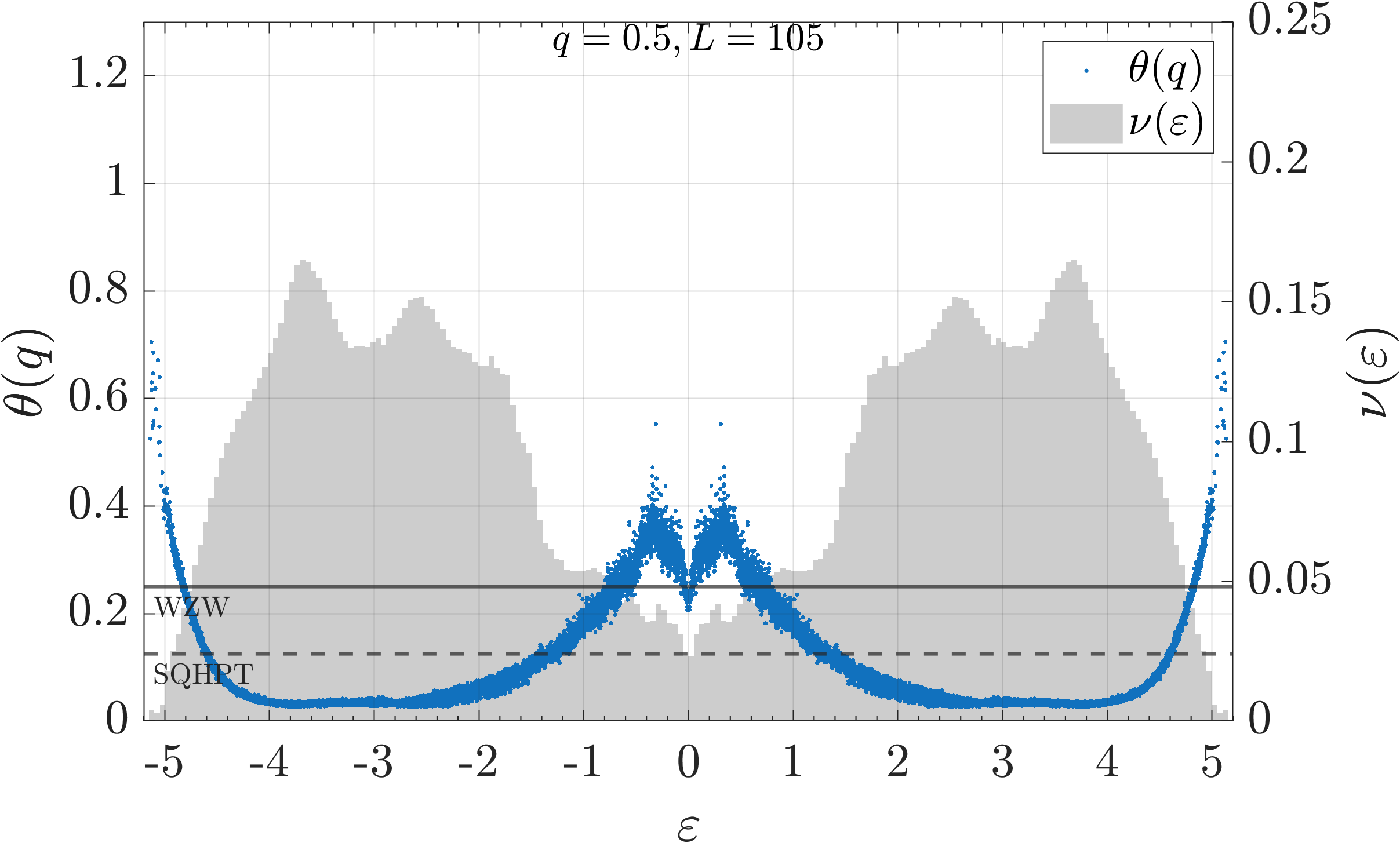}
  }
  \caption{\label{fig:bm-weak-dis}
    Effects of chiral ``twist'' disorder [Eq.~(\ref{TwistDirt})] on the chiral Bistritzer-MacDonald model for
    TBLG [Eq.~(\ref{TBLG})]. 
    This is a special case of the 2D continuum model without the flat band, whereby spectral gaps
    arise from the (periodic, non-random) moir\'e potential.
    We incorporate weak uncorrelated randomness in each component of the latter. 
    The model is tuned to the first magic angle $\alpha \simeq 0.586$.
    As in Figs.~\ref{fig:ci_conti_theta} and \ref{fig:ci_conti_flat_theta}, we plot the multifractal curvature $\theta_q (\varepsilon)$
    and density of states $\nu(\e)$ versus energy $\e$, for different disorder strengths
    (a) $\lambda = 0.5$ and (b) $\lambda = 2 \pi$.
    As for the case of the continuum coupled to the flat band Fig.~\ref{fig:ci_conti_flat_theta},
    the initial localizing effects of weak disorder appear to be \emph{suppressed} by stronger disorder, ``healing'' 
    the surface. 
    The moir\'e parameters (see Fig.~\ref{fig:BM_Lattice} and associated discussion in text) are $L = 105$ and $\xi = \pi$.}
\end{figure}

\begin{figure}[b!]
  \centering
  \subfigure[]{
    \includegraphics[width=0.4\textwidth]{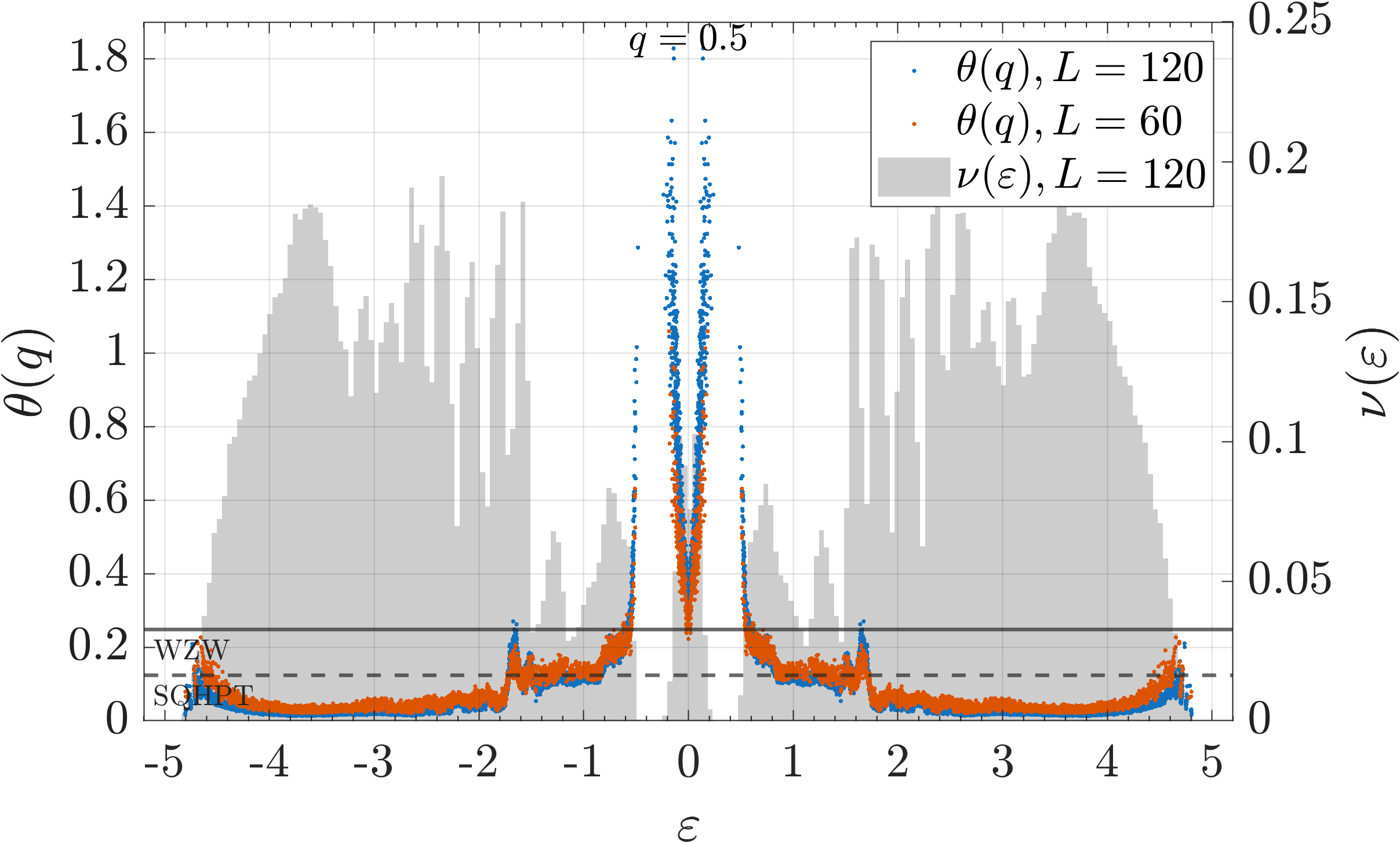}
  }
  \subfigure[]{
    \includegraphics[width=0.4\textwidth]{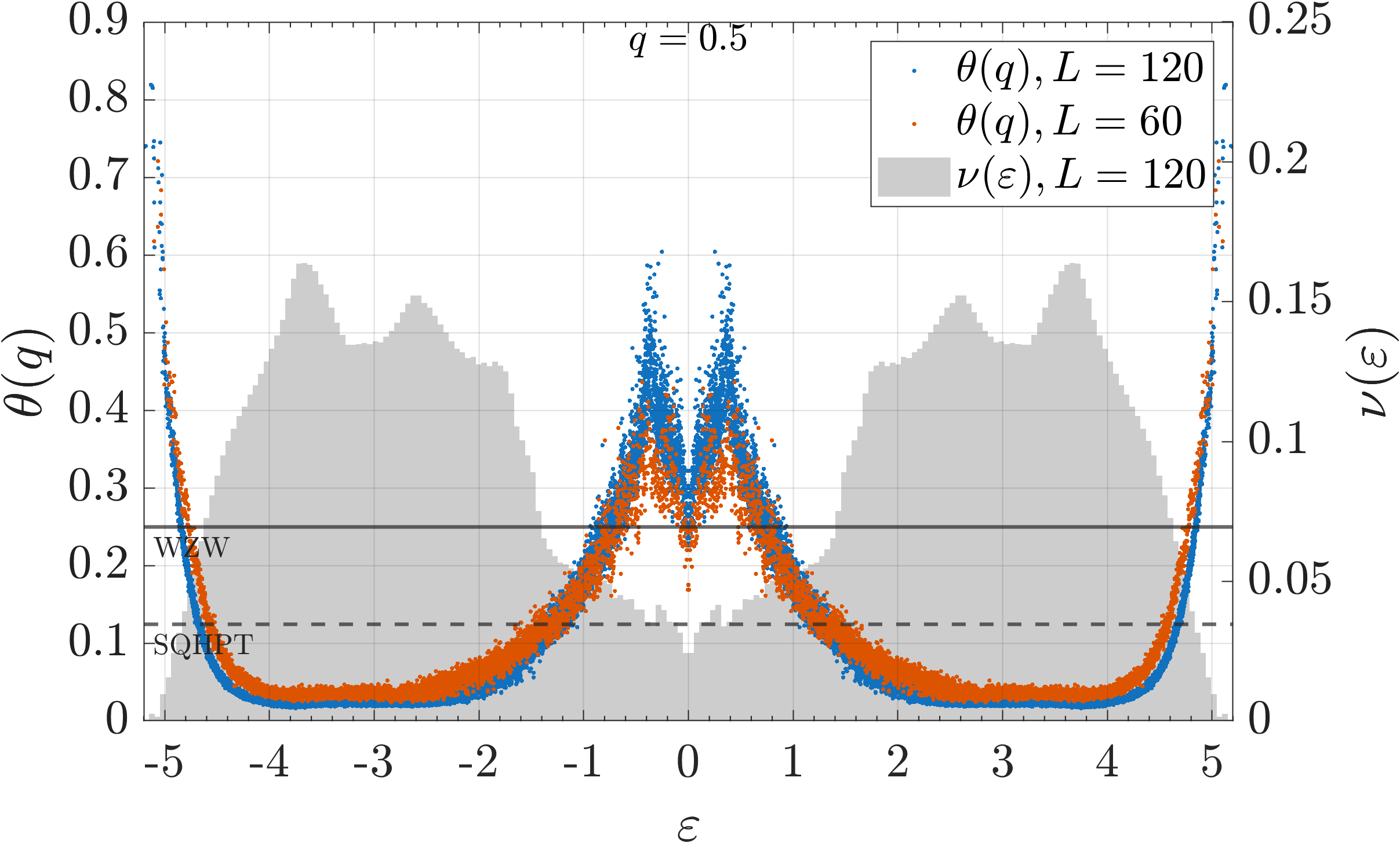}
  }
  \caption{\label{fig:bm-strong-dis}
  Same as Fig.~\ref{fig:bm-weak-dis}, but with multifractal spectra shown for two different system sizes
  $L = 90$, box sizes $b = (3, 5, 6)$ and $L = 120$, $b = (4, 6, 8, 10)$.
  (a) $\lambda_a = 0.5$. 
  (b) $\lambda_a = 2 \pi$.  
  The correlation length is  $\xi = \pi$.}
\end{figure}

\subsection{CI continuum model, no TPE}

Next we examine the 2D continuum Dirac theory defined by Eq.~(\ref{hx2D}), previously studied in Ref.~\cite{Ghorashi18}. 
We use exact diagonalization to solve for the eigenstates in momentum space (avoiding fermion doubling of the continuum
Dirac equation). We compare results for the surface without and with the flat band, Eq.~(\ref{HDiracTPE}). 
Disorder is incorporated via Eq.~(\ref{eq:ci_continu_a}).

Results without the flat band are exhibited in Figs.~\ref{fig:ci_continu} and \ref{fig:ci_conti_theta}. 
Fig.~\ref{fig:ci_continu}(a) compares the level statistics 
\cite{RMRev}
to the Gaussian orthogonal (GOE) and Gaussian unitary (GUE) ensembles;
the results are consistent with the former. This is expected from standard arguments, whereby finite-energy states of class-CI Hamiltonians reside in the orthogonal class AI \cite{Ghorashi18,Evers08}. The global density of states is shown in Fig.~\ref{fig:ci_continu}(b), and exhibits the low-energy cusp expected for class CI, see Ref.~\cite{Ghorashi18} for a quantitative comparison to the analytical prediction \cite{Nersesyan1994}. 
Despite the GOE statistics, Fig.~\ref{fig:ci_conti_theta} demonstrates that \emph{most} surface states converge to $\theta \sim 1/8$
with increasing disorder (except those far above the energy cutoff, in the high-energy Lifshitz tails), 
consistent with class-C spin quantum Hall plateau transition (SQHPT). This is the SWQC scenario \cite{Ghorashi18,Karcher21}. The orthogonal class is instead expected to Anderson localize in 2D.

Fig.~\ref{fig:ci_conti_flat_theta} shows the effects of disorder on the hybridized continuum 2D surface-state--flat-band system. 
Weak disorder preserves the spectral gaps opened by hybridization, and localizes states close to the gap edges. 
This suggests the onset of the TPE. Different from the lattice results shown in Figs.~\ref{fig:ci_lattice_theta_dis}(b) 
and \ref{fig:ci_lattice_theta_n}(b), however, increasing disorder for the continuum system appears to \emph{heal} the surface.
I.e., spectral gaps are washed out, and the multifractal spectrum re-acquires the SWQC demonstrated in Fig.~\ref{fig:ci_conti_theta} without the flat band
hybridization. 

The key takeaway is that the \emph{continuum and lattice models behave oppositely with increasing disorder}. 
The lattice results agree with the 
picture articulated
in Ref.~\cite{UFO24},
which correlated surface-state fragility with the hybridization to trivial degrees of freedom; 
the combined system is expected to exhibit the TPE. 
By contrast, the continuum results do \emph{not} show a robust TPE. 
Unlike the lattice-scale (``fragmenting potential'') device used in Ref.~\cite{UFO24} to localize class-AIII surface states, it was argued in this reference that the TPE would destabilize \emph{both} bulk lattice and 2D continuum-model surface states. The surprising finding that only the lattice model shows the expected behavior is a second main result of this paper. 

Examining Fig.~\ref{fig:ci_conti_flat_theta}, one could argue that the results are not unexpected. The clean flat band, surface-state hybridized system exhibits energy gaps that 
disappear
when the disorder becomes sufficiently strong so as to ``smear out'' the position of the flat band. Nevertheless, Fig.~\ref{fig:ci_conti_flat_theta}(c) shows that the outcome for strong disorder is essentially identical to the isolated surface [Fig.~\ref{fig:ci_conti_theta}(c)]. 

At this point it is worth reiterating that the results for the isolated surface \cite{Ghorashi18} are anomalous and \emph{not} expected. According to the standard logic, all finite-energy states should reside in the orthogonal class AI. For disorder not too strong and in finite volume, 2D wave functions will appear critical with an approximately parabolic multifractal spectrum. The parameter $\theta(\e)$ is determined by dimensionless conductance $G(\e)$ of the states \cite{Wegner80},
\begin{align}
    \theta(\e) = \frac{1}{2 \pi^2 G(\e)}.
\end{align}
Ignoring slow logarithmic renormalization effects (weak localization), the 
dimensionless finite-energy $G(\e)$ should vary inversely with the disorder strength $\lambda_a$, 
so that $\theta(\e) \sim \lambda_a$.
Instead, both Figs.~\ref{fig:ci_conti_theta} and \ref{fig:ci_conti_flat_theta} show convergence
to the universal SQHPT value $\theta(\e) \sim 1/8$ with increasing disorder.
The conclusion is that even the flat-band hybridized surface theory seems to show robust SWQC \cite{Ghorashi18,Karcher21}, at odds with the lattice-model results obtained above.

\subsection{Bistritzer-MacDonald model}

As a final point of comparison, we consider the chiral BM model for TBLG [Eq.~(\ref{TBLG})], subject to 
the chiral disorder in Eq.~(\ref{TwistDirt}). As explained in Sec.~\ref{sec:cTBLG}, the chiral BM model 
is a special case of the 2D class-CI Dirac continuum model defined by Eq.~(\ref{hx2D}), with
four of the six vector-potential components encoding the interlayer (pure $AB$) tunneling due to the moir\'e 
potential, Eq.~(\ref{moirepot}). At the first magic angle, the chiral BM model has spectral gaps between 
the zero-energy flat bands and the rest of the spectrum \cite{BM11,Tarnopolosky19}. 
This already indicates the absence of a spectral flow principle for class-CI surface states. 

In this case, we consider the incorporation of ``twist'' disorder 
\cite{wilson2020disorder,Padhi20,shavit2023strain,nakatsuji2022moire,guerrero2025disorderinduced,sanjuanciepielewski2025transport,Queiroz25,
uri2020mapping,kapfer2023programming} in the form of random white-noise additions to the moir\'e components,
Eq.~(\ref{TwistDirt}). Results are shown in Figs.~\ref{fig:bm-weak-dis} and \ref{fig:bm-strong-dis}. 
As in the case of the continuum model without the moir\'e potential, but with hybridization to the continuum
flat band (Fig.~\ref{fig:ci_conti_flat_theta}), the numerical results for the dirty chiral BM model 
show that the energy gaps are not stable to disorder. 
Strongly localized states that appear near the gap edges for weak disorder are  ``washed out'' by stronger disorder. 
The overall trend of the states, averaged over energy, is a \emph{suppression} of the extent of Anderson localization
with increasing disorder, see Fig.~\ref{fig:theta}. 

These results obtained here for ``twist'' dirt in the chiral BM model are very similar to the continuum case without the moir\'e
potential: localization is suppressed by increasing disorder, which also fills in the spectral gaps. 
For the disorder strengths studied in this case, we do not observe the class-C SWQC for the BM model. This is because the model remains dominated by the periodic BM potential at larger energies for the system sizes that we can access here. We note that although the disorder studied here is chiral, it is different from the fine-tuned
disorder discussed in Ref.~\cite{Queiroz25} that preserves the flatness of the zero-energy bands at the first magic angle.


\section{Open questions \label{sec:oq}}

As we have already summarized the main problems explored and resulting findings in the Introduction, 
here we close with a set of important open questions. 

\begin{enumerate}
\item{What is the physical picture behind the trivialization of surface states in class-CI bulk topological lattice models? For class AIII, a Berry curvature mechanism was uncovered in Ref.~\cite{UFO24}, reviewed in Sec.~\ref{sec:SWQC}. This cannot apply to class CI due to the $T^2 = +1$ time-reversal symmetry in the latter.
One possibility is quantum geometry \cite{Provost80,torma2023essay,QGRev1,QGRev2}, although for the case studied here the surface-trivial band hybridization does \emph{not} open up a spectral gap, see Fig.~\ref{fig:lattice_flat_spec}(a). 
Moreover, the absence of a robust TPE for the continuum model studied here (with a spectral hybridization gap in the clean limit) suggests that 
one must confront the quantum geometry of bands that \emph{interpolate} between 2D to 3D in the slab geometry. 
}
\item{A second key result of this work is the divergence between lattice and continuum surface models regarding the fragility of class-CI surface states. This points to a fundamental incompleteness of the 2D continuum Dirac equation for describing topological surface states. In particular, the TPE (introduced in Ref.~\cite{UFO24} but explored for the first time numerically here) gives conflicting results for bulk lattice versus continuum surface calculations. Are there important implications for twisted moir\'e materials such as TBLG? This is an especially interesting, but potentially fraught question, given the \emph{lattice-scale quasiperiodicity} (i.e., non-translational invariance) of twisted multilayers.
}
\item{Finally, another persistent mystery is the \emph{delocalization mechanism} for class-CI surface states at finite energy, i.e.\ the spectrum-wide quantum criticality scenario. For class AIII, the observation of critical, integer-quantum-Hall-plateau states at all nonzero surface-state energies is understood as the result of a statistical symmetry, tuning between surface-state topological Anderson insulators that are nucleated by Berry curvature from the surface-bulk interface or TPE \cite{UFO24}. From a symmetry-class perspective, this is natural because finite-energy states of any class-AIII system are expected to reside in the unitary class A, which also hosts the quantum Hall effect \cite{Evers08}.

By contrast, finite-energy states of a class-CI system (topological or trivial, bulk or surface, lattice or continuum) are expected to reside in the orthogonal class AI \cite{Ghorashi18}, a class that is believed to \emph{always localize} in two spatial dimensions \cite{Evers08}. In Ref.~\cite{Ghorashi18}, it was argued that class-C might alternatively be realized for finite-energy surface states of a class-CI bulk, based in part on a symmetry analysis of replicated nonlinear sigma model target manifolds. The numerical results for SWQC obtained here and in \cite{Ghorashi18,Karcher21} are consistent with the spin quantum Hall plateau transition in class C, but no physical picture has yet emerged to convincingly explain this observation. An alternative scenario could be an (at present unknown) delocalization mechanism for class AI that happens to give a very similar multifractal spectrum as the class-C plateau transition. 
}
\item{We note that continuum \emph{bulk} theories are sometimes used to model topological materials. 
The results obtained here and in Ref.~\cite{UFO24} imply that bulk \emph{Dirac} models are problematic,
as 2D Dirac surface states in such models typically co-exist (and never merge with) the bulk up to arbitrarily large energies. On the other hand, surface states span a finite range and merge with the bulk in some faithful 
non-relativistic continuum models, e.g.\ for 3-He-B \cite{He3Rev}.}  
\end{enumerate}

\acknowledgments
We thank L.\ Trifunovic, P.\ Brouwer, and R.\ Queiroz for helpful discussions. This work was supported by the Welch Foundation Grant No.~C-1809. This work was supported in part by the Big-Data Private-Cloud Research Cyberinfrastructure MRI-award funded by NSF under grant CNS-1338099 and by Rice University's Center for Research Computing (CRC).


\appendix


\section{Parameters for Fig.~\ref{fig:theta} \label{sec:fig1params}}

The data points presented in Fig.~\ref{fig:theta} of the Introduction are obtained for the five 
different models studied in this work. The results in this summary figure involve averaging
multifractal curvature data over finite energy windows in each model for different disorder strengths. 
The precise specification of all parameters used to obtain Fig.~\ref{fig:theta} are as follows,
\begin{enumerate}
  \item CI lattice model. $W_1=0.4, 0.7, 0.85, 1$. Parameters: $N_y = 72$, $N_z = 10$, $\mu = - 2$,
      $\Delta_1 = 1$, $\Delta_2 = 1$, $m_c = 1$. 
      Energy window $(0.1, \infty)$.
  \item CI lattice flat band. $W_1=0.4, 0.7, 1, 1.2$. Parameters: $\varepsilon_c = 0.4$, $\gamma = 0.5$,
      $N_y = 72$, $N_z = 10$, $\mu = - 2$, $\Delta_1 = 1$, $\Delta_2 = 1$, $m_c = 1$. 
      Energy window $(0.1, \infty)$.
  \item CI continuum model. $W=0.5, 1, \pi, 2\pi$. Parameters: $L = 97$, $\xi = \pi$, $b = (4, 6, 8)$. 
      Energy window $(0.1, 0.4)$.
  \item CI continuum model with flat band. $W=0.2, 1, \pi, 2\pi$. Parameters: $\varepsilon_c = 0.4$,
      $\lambda_c = 0.2$, $\gamma_0 = 0.35$, $\lambda_g = 0.2$, $L = 97$, $\xi = \pi$, $b = (4, 6, 8)$. 
      Energy window $(0.1, 0.6)$.
  \item BM model. $W=0.5, \pi, 2\pi, 7$. Parameters: $M_1 = 5$, $M_2 = 7$, $L = 105$. 
        Energy window $(0.1, 1.5)$.
\end{enumerate}


\begin{thebibliography}{99}
\bibitem{DIIInote}
    In addition to localizable classes CI, CII, and AIII, 
    class DIII, which includes $^3$He-$B$ \cite{Schnyder08}, is non-localizable (localizable) for odd (even)
    winding numbers in three dimensions \cite{UFO24}.
\bibitem{moiresimnote}
    In the case of twisted bilayer, the chiral limit is equivalent to a class-CI surface theory.
    Incorporating $AA$ interlayer hopping but preserving approximate particle-hole symmetry 
    (the model with ``stable topology'' \cite{Song19,Song21}) can be viewed as
    class-CI surface states with time-reversal symmetry breaking (class C) that does not open a gap.
    Breaking both chiral and particle-hole symmetries in twisted bilayer 
    (``fragile topology'' \cite{Po18,Po19,Ahn19})
    gives a Dirac Hamiltonian in class A that is equivalent to a non-topological 
    spinless graphene monolayer with intervalley coupling.
\end{thebibliography}
\end{document}